\documentclass[10pt, twocolumn]{revtex4-1}

\usepackage{amsmath}
\usepackage[dvipdfmx]{graphicx}
\usepackage{amssymb}

\usepackage{color}
\usepackage{ulem}
\newcommand{\mf}[2]{\textcolor{blue}{#1}\textcolor{blue}{\sout{\textcolor{red}{#2}}}}

\begin{document}
\title{
Robustness against additional noise in cellular information transmission
}
\author{Takehiro Tottori}
\affiliation{Department of Mathematical Informatics, Graduate School of Information Science and Technology, University of Tokyo, Tokyo, Japan}
\author{Masashi Fujii}
\email{m-fujii0123@hiroshima-u.ac.jp}
\affiliation{Department of Integrated Sciences for Life, Graduate School of Integrated Sciences for Life, Hiroshima University, Hiroshima, Japan}
\author{Shinya Kuroda}
\affiliation{Department of Biological Sciences, Graduate School of Science, University of Tokyo, Tokyo, Japan}


\begin{abstract}
Fluctuations in intracellular reactions (intrinsic noise) reduce the information transmitted from an extracellular input to a cellular response. However, recent studies have demonstrated that the decrease in the transmitted information with respect to extracellular input fluctuations (extrinsic noise) is smaller when the intrinsic noise is larger. Therefore, it has been suggested that robustness against extrinsic noise increases with the level of the intrinsic noise. We call this phenomenon intrinsic noise-induced robustness (INIR). As previous studies on this phenomenon have focused on complex biochemical reactions, the relation between INIR and the input--output of a system is unclear. Moreover, the mechanism of INIR remains elusive. In this paper, we address these questions by analyzing simple models. We first analyze a model in which the input--output relation is linear. We show that the robustness against extrinsic noise increases with the intrinsic noise, confirming the INIR phenomenon. Moreover, the robustness against the extrinsic noise is more strongly dependent on the intrinsic noise when the variance of the intrinsic noise is larger than that of the input distribution. Next, we analyze a threshold model in which the output depends on whether the input exceeds the threshold. When the threshold is equal to the mean of the input, INIR is realized, but when the threshold is much larger than the mean, the threshold model exhibits stochastic resonance, and INIR is not always apparent. The robustness against extrinsic noise and the transmitted information can be traded off against one another in the linear model and the threshold model without stochastic resonance, whereas they can be simultaneously increased in the threshold model with stochastic resonance.
\end{abstract}
\maketitle

\section{Introduction} 
The transmission of cellular information plays an important role in various biological functions. In this process, environmental information is encoded to extracellular molecules, which generate a cellular response (Fig. \ref{fig-Schematic diagram of cellular information transmission}(a)). However, in many cases, extracellular molecules exhibit considerable fluctuations (extrinsic noise). For example, in developmental processes, cells detect their positional information through morphogen molecules. However, the concentration of morphogen molecules fluctuates because of the stochasticity of the diffusions \cite{Houchmandzadeh2002}. Another example is chemotaxis, in which the reaction between extracellular molecules and receptors largely fluctuates because of the stochasticity of the reaction \cite{Ueda2007,Miyanaga2007}. Regardless of such noisy conditions, cells often respond to their environments robustly, and this response mechanism has attracted considerable attention \cite{Libby2007,Andrews2007,Tkacik2008a,Tkacik2008b,Morishita2008,Tostevin2009,Kobayashi2010,Mugler2013,Hasegawa2018,Monti2018}.

Intracellular reactions also exhibit significant fluctuations because of the stochasticity of the reactions \cite{Elowitz2002,Bar-Even2006,Newman2006,Taniguchi2010}. The fluctuations of intracellular reactions (intrinsic noise) generally reduce the information transmitted from the environment to the cellular response \cite{Gregor2007,Cheong2011,Uda2013,Selimkhanov2014}, and so intrinsic noise is thought to be disadvantageous for cellular information transmission. However, are there no advantageous effects of intrinsic noise?

Recent studies simulating biochemical reactions have demonstrated that the transmitted information barely decreases with respect to extrinsic noise when the intrinsic noise is large \cite{Koumura2014,Fujii2017,Tottori2019}. In other words, robustness against the extrinsic noise increases with the level of the intrinsic noise. It has therefore been suggested that robustness against extrinsic noise is one of the advantageous characteristics of intrinsic noise \cite{Koumura2014,Fujii2017,Tottori2019}. In this paper, we call this phenomenon intrinsic noise-induced robustness (INIR).

Previous studies on this mechanism have focused on specific biochemical reactions \cite{Koumura2014,Fujii2017,Tottori2019}; hence, their models were too complicated to clarify the relation between INIR and the input--output of a system. Moreover, the mechanism of INIR remains elusive. In this paper, we address these issues by analyzing simple models. 

This paper is organized as follows. In Sec. \ref{section-Linear model}, we analyze a model in which the input--output relation is linear. This model is an extension of the Gaussian channel, which is a standard model in information theory \cite{Cover}. We show that the decrease in transmitted information with respect to extrinsic noise is smaller when the intrinsic noise is larger. Therefore, even this simple model reproduces the INIR phenomenon. Moreover, we demonstrate that the intrinsic noise dependency of the robustness against the extrinsic noise becomes stronger when the variance of the intrinsic noise is larger than that of the input distribution. 

In Sec. \ref{section-Threshold model}, we analyze a threshold model in which the output depends on whether the input exceeds the threshold. When the threshold is equal to the mean of the input, INIR is realized in the threshold model (Sec. \ref{subsection-theta0}). In contrast, when the threshold is much larger than the mean of the input, the threshold model exhibits stochastic resonance (SR) (Sec. \ref{subsection-SR}). Under the SR phenomenon, noise acts to increase the level of transmitted information \cite{Gammaitoni1998}, and it is experimentally verified that SR appears in biological systems such as neurons \cite{Douglass1993,Levin1996,Russell1999,Gluckman1996,Stacey2000,Stacey2001,Srebro1999,Mori2002,Kitajo2007,Stufflebeam2000,Tanaka2008,Ward2010}. Interestingly, when the threshold model exhibits SR, INIR is not always realized. However, the threshold model with SR can increase both robustness against extrinsic noise and the transmitted information simultaneously, which is impossible for the linear model and the threshold model without SR. In other words, SR solves the trade-off between robustness against extrinsic noise and the transmitted information.

In Sec. \ref{section-Mechanism}, we explain why INIR is realized in the linear model and in the threshold model without SR, but is not always realized in the threshold model with SR. In Sec. \ref{section-Discussion}, we discuss the biological significance of our results and the relevance of other studies.

\section{Linear model\label{section-Linear model}}

\begin{figure}[btp]
\begin{center}
	\includegraphics[width=75mm]{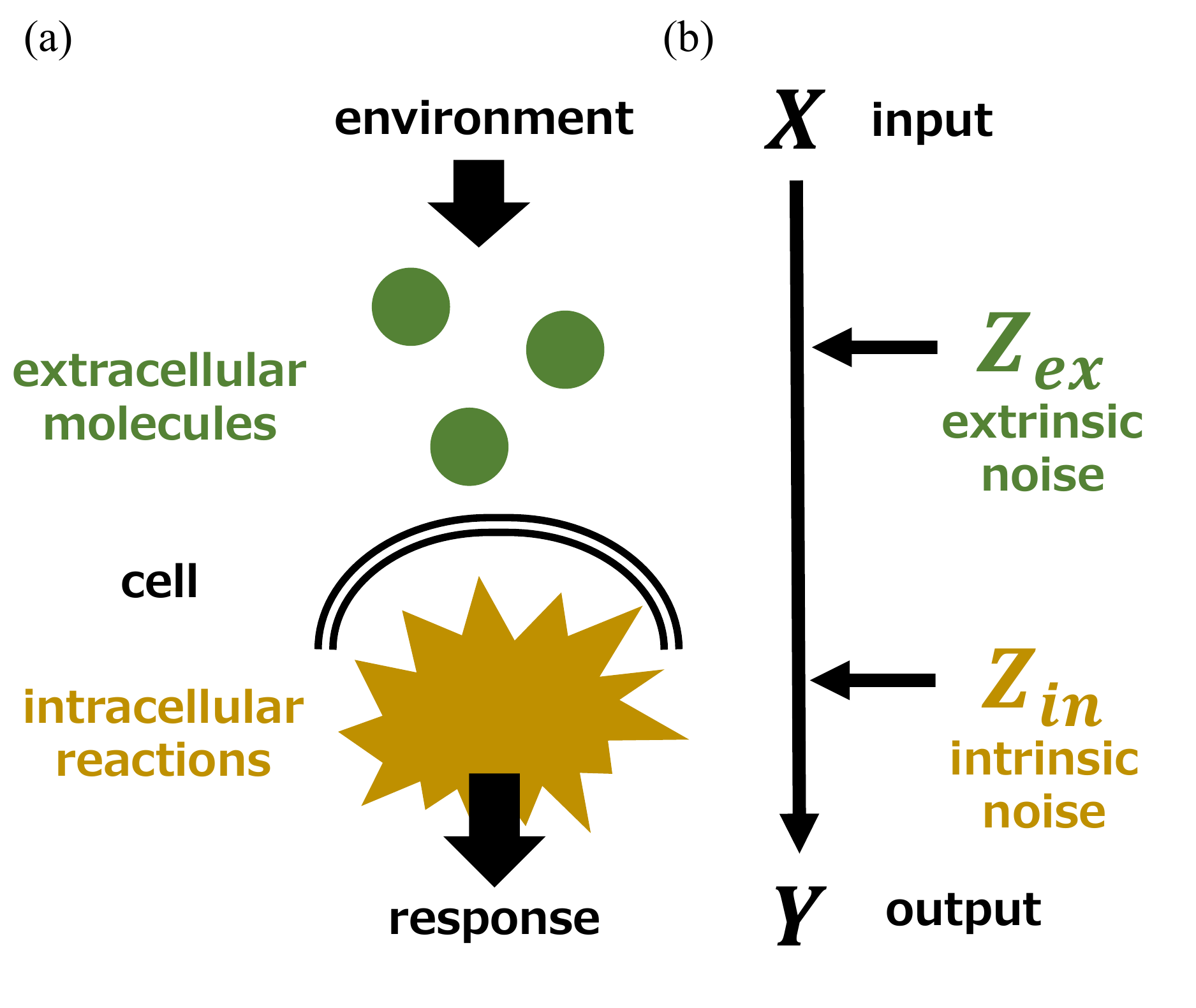}\\[-1pc]
	 \caption{
	Schematic diagram of cellular information transmission.
	(a) 
	Environmental information is encoded to extracellular molecules. Cells detect extracellular molecules and determine proper responses via intracellular reactions.
	(b)
	The environment and the cellular responses correspond to input $X$ and output $Y$, respectively. Furthermore, fluctuations of extracellular molecules and intracellular reactions correspond to extrinsic noise $Z_{ex}$ and intrinsic noise $Z_{in}$, respectively.
	}
	\label{fig-Schematic diagram of cellular information transmission}
\end{center}
\end{figure}

We consider a channel composed of the input $X$, extrinsic noise $Z_{ex}$, intrinsic noise $Z_{in}$, and output $Y$ (Fig. \ref{fig-Schematic diagram of cellular information transmission}). The input $X$ corresponds to environmental information detected by the cells, and the output $Y$ corresponds to the cellular response against the environment. Furthermore, the extrinsic noise $Z_{ex}$ corresponds to the fluctuation of extracellular molecules, and the intrinsic noise $Z_{in}$ corresponds to the fluctuation of intracellular reactions. For example, in developmental processes, $X$ corresponds to the concentration of morphogen molecules, which encodes the positional information, and $Y$ corresponds to the concentration of gene expressions \cite{Tkacik2008a,Tkacik2008b}.  $Z_{ex}$ corresponds to the fluctuation of morphogen molecules because of the stochasticity of diffusions \cite{Houchmandzadeh2002}, and $Z_{in}$ corresponds to the fluctuation of gene expressions caused by the stochasticity of intracellular reactions \cite{Gregor2007,Tkacik2008a,Tkacik2008b}. 

Here, the input $X$, extrinsic noise $Z_{ex}$, and intrinsic noise $Z_{in}$ are assumed to obey Gaussian distributions $N(0,1)$, $N(0,\sigma_{ex}^2)$, and $N(0,\sigma_{in}^2)$, respectively, where $\sigma_{ex}^2$ is the variance of the extrinsic noise, and $\sigma_{in}^2$ is the variance of the intrinsic noise. Furthermore, the output $Y$ is given by the following:

\begin{align}
	Y&=X+Z_{ex}+Z_{in}
	\label{Linear response}
\end{align}
This model is an extension of the Gaussian channel, which is a standard model in information theory \cite{Cover}. In this paper, we call this the ``linear model.''

To quantify the information transmitted from the input $X$ to the output $Y$, we use the mutual information between $X$ and $Y$, $I(X;Y)$ \cite{Cover}. Higher values of $I(X;Y)$ denote that a large amount of information is transmitted from $X$ to $Y$. When $X$ and $Y$ are continuous random variables, $I(X;Y)$ is given by \cite{Cover}:

\begin{align}
	I(X;Y)&=\int_{Y}\int_{X}p(x,y)\log_{2}\frac{p(x,y)}{p(x)p(y)}dxdy
\end{align}
where $p(x,y)$ is the joint probability density function of $X$ and $Y$, and $p(x)$ and $p(y)$ are the marginal probability density functions of $X$ and $Y$, respectively. In the linear model, $I(X;Y)$ can be derived analytically as \cite{Cover}:

\begin{align}
	I(X;Y)&=\frac{1}{2}\log_{2}\left(1+\frac{1}{\sigma_{ex}^2+\sigma_{in}^2}\right)
	\label{LRM-MI}
\end{align}

\begin{figure}[btp]
\begin{center}
	(a)\hspace{-3.0mm}
	\begin{minipage}[t][][b]{40mm}
		\includegraphics[width=40mm,height=35mm]{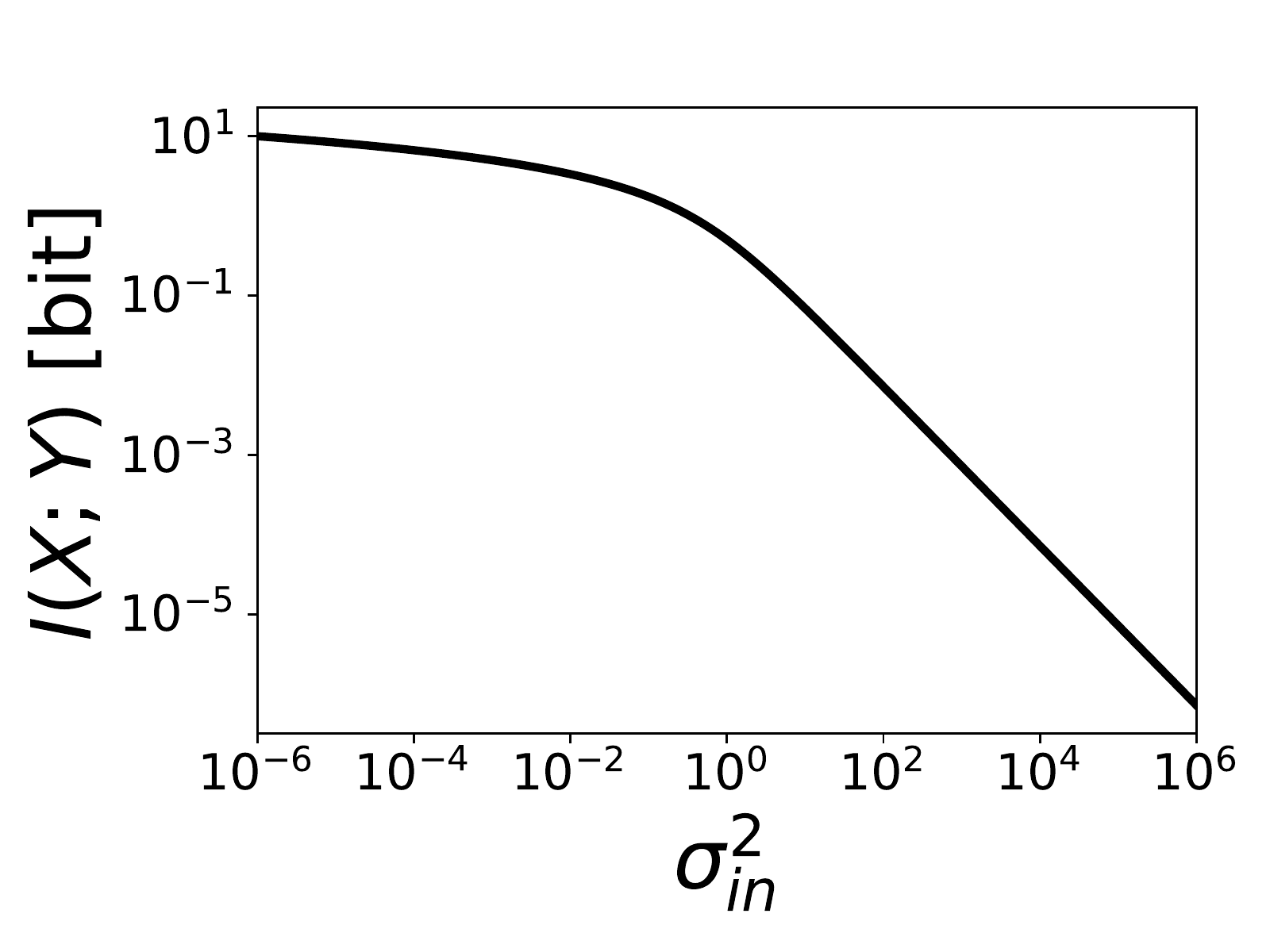}
	\end{minipage}
	(b)\hspace{-3.0mm}
	\begin{minipage}[t][][b]{40mm}
		\includegraphics[width=40mm,height=35mm]{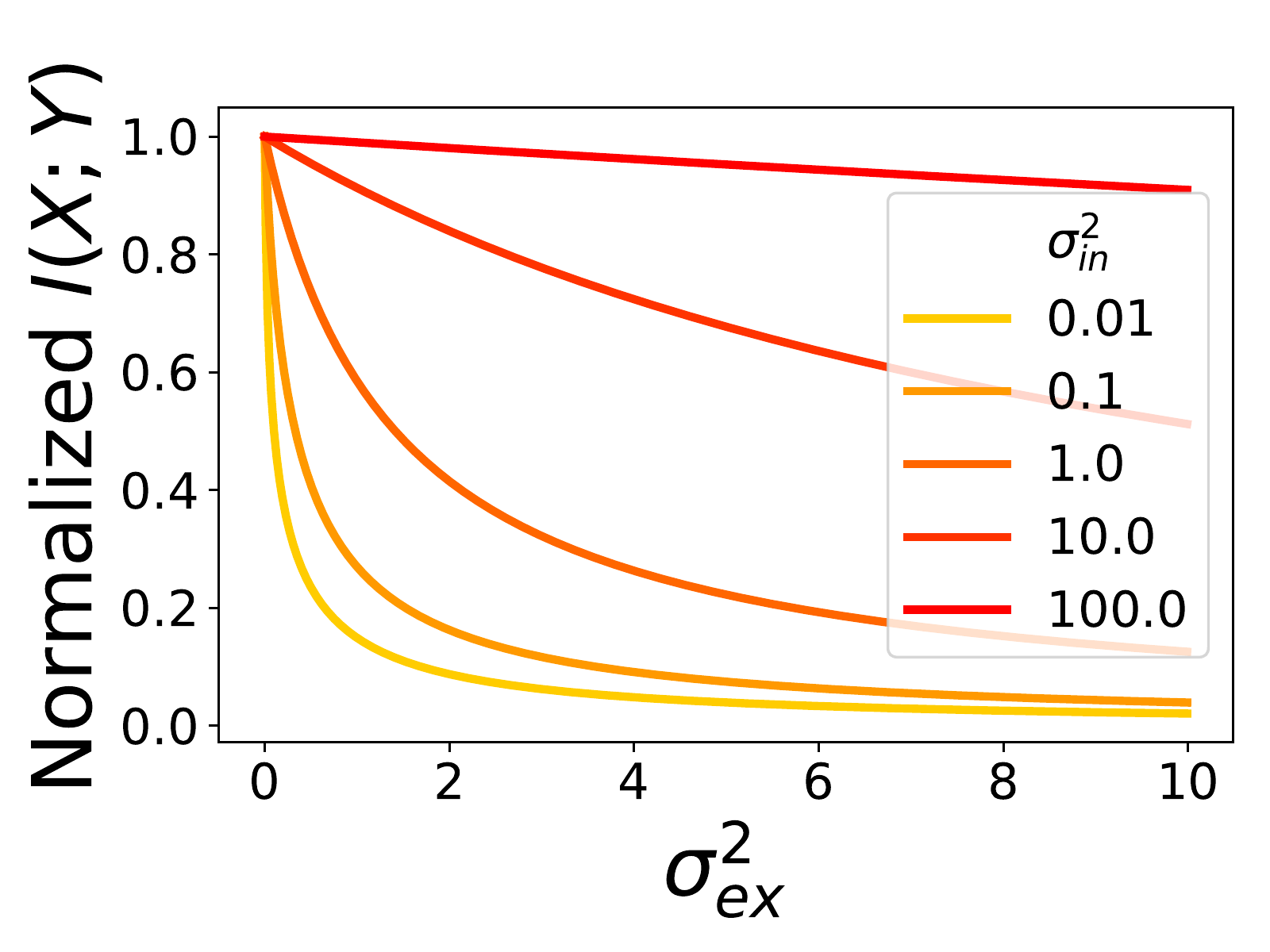}
	\end{minipage}\\
	(c)\hspace{-3.0mm}
	\begin{minipage}[t][][b]{40mm}
		\includegraphics[width=40mm,height=35mm]{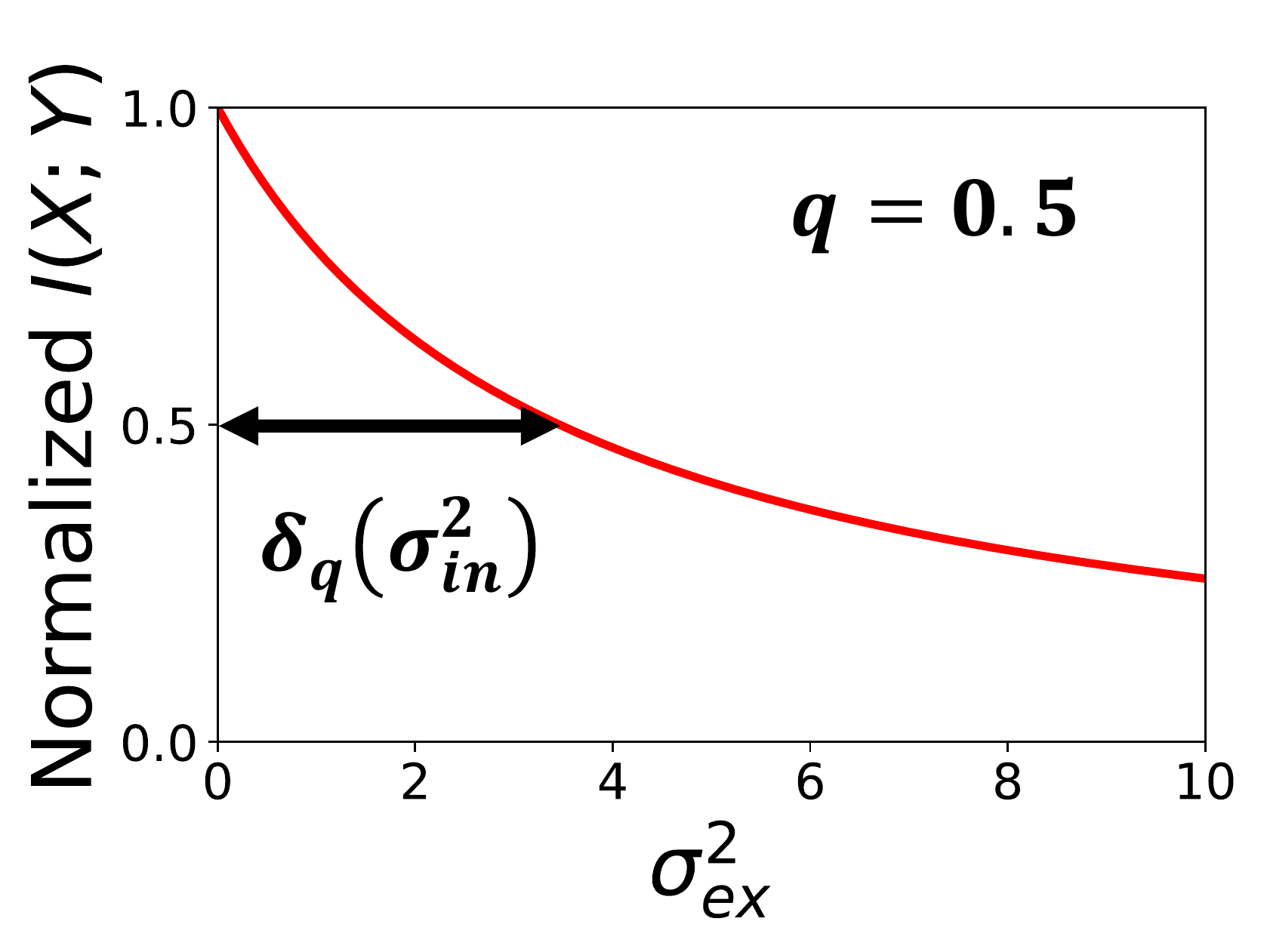}
	\end{minipage}
	(d)\hspace{-3.0mm}
	\begin{minipage}[t][][b]{40mm}
		\includegraphics[width=40mm,height=35mm]{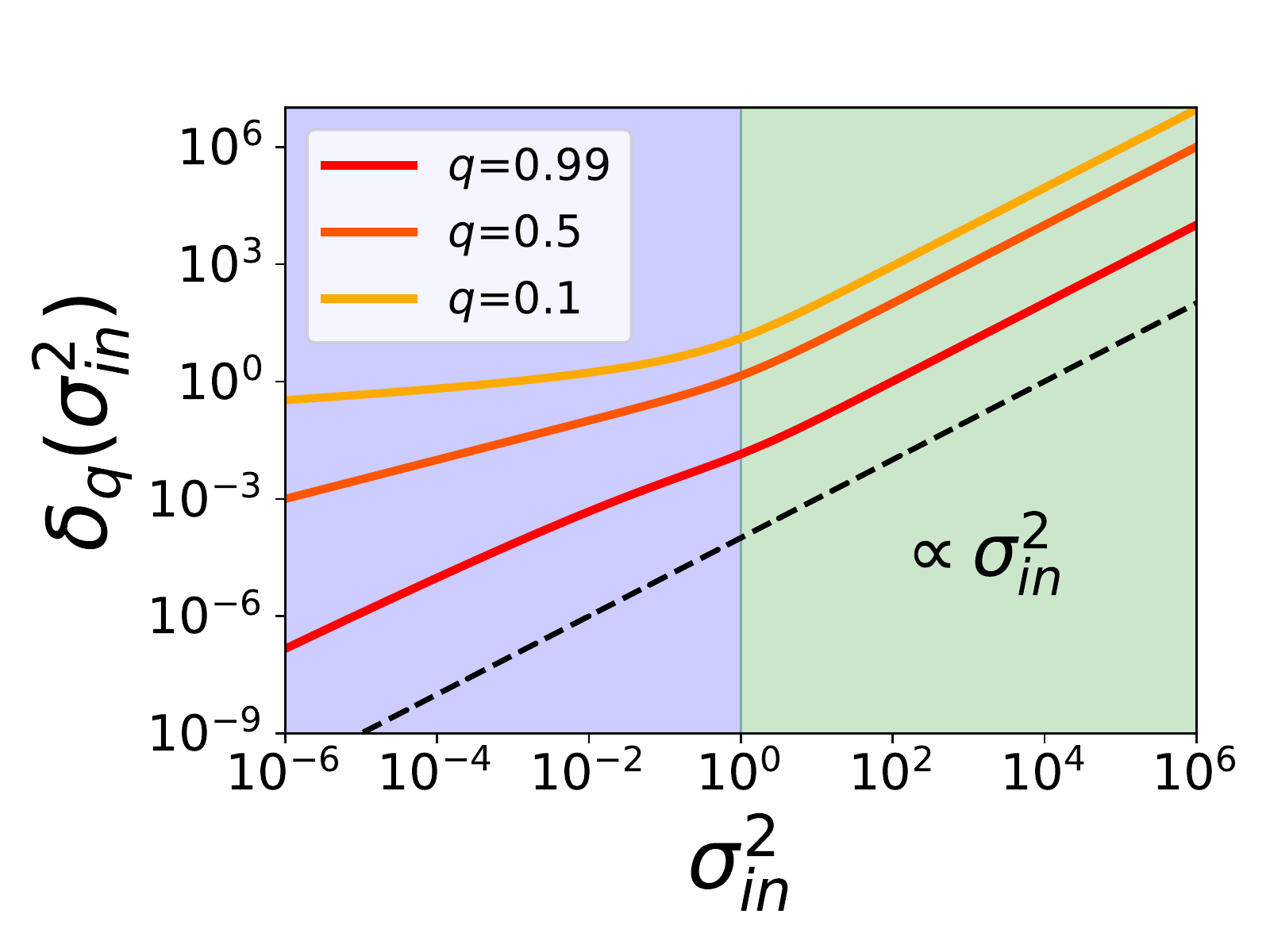}
	\end{minipage}
	 \caption{
	 Linear model. 
	 (a) 
	 Mutual information $I(X;Y)$ against intrinsic noise $\sigma_{in}^2$ at $\sigma_{ex}^2=0$. 
	 (b) 
	 Mutual information $I(X;Y)$ against extrinsic noise $\sigma_{ex}^2$. The mutual information $I(X;Y)$ is normalized at $\sigma_{ex}^2=0$.
	 (c)
	 Definition of the robustness function $\delta_{q}(\sigma_{in}^2)$.
	 (d)
	 Intrinsic noise dependency of the robustness function $\delta_{q}(\sigma_{in}^2)$. While $\delta_{q}(\sigma_{in}^2)\propto\sigma_{in}^2$ when $\sigma_{in}^2>1$ (green), $\delta_{q}(\sigma_{in}^2)\propto\sigma_{in}^{2q}$ when $\sigma_{in}^2<1$ (blue). 
	 }
	\label{linear model}
\end{center}
\end{figure}

The mutual information $I(X;Y)$ decreases with the intrinsic noise $\sigma_{in}^2$ when $\sigma_{ex}^2=0$ (Fig. \ref{linear model}(a)). However, the decrease in $I(X;Y)$ with respect to extrinsic noise $\sigma_{ex}^2$ is reduced at higher levels of intrinsic noise $\sigma_{in}^2$ (Fig. \ref{linear model}(b)). In other words, the degree of robustness against extrinsic noise increases with the level of intrinsic noise. Therefore, INIR is realized in the linear model.

To investigate the robustness against extrinsic noise in more detail, we define the robustness function $\delta_{q}(\sigma_{in}^2)$ as follows:

\begin{align}
	\delta_{q}(\sigma_{in}^2):=\left\{\sigma_{ex}^{*2}\left|\frac{I(X;Y)|_{\sigma_{ex}^2=\sigma_{ex}^{*2}}}{I(X;Y)|_{\sigma_{ex}^2=0}}=q,\ \ \ q\in(0,1)\right.\right\}
	\label{the definition of the robustness function}
\end{align}

Therefore, the robustness function $\delta_{q}(\sigma_{in}^2)$ indicates the extrinsic noise $\sigma_{ex}^2$ when the normalized mutual information $I(X;Y)/I(X;Y)|_{\sigma_{ex}^2=0}$ decreases to $q$ for $q\in(0,1)$ (Fig. \ref{linear model}(c)). For example, when $q=0.5$, $\delta_{q}(\sigma_{in}^2)$ indicates the extrinsic noise $\sigma_{ex}^2$ when the mutual information $I(X;Y)$ has decreased by 50\%; when $q=0.99$, the robustness function $\delta_{q}(\sigma_{in}^2)$ indicates the extrinsic noise $\sigma_{ex}^2$ when the mutual information $I(X;Y)$ has decreased by 1\%. Higher values of $\delta_{q}(\sigma_{in}^2)$ indicate higher robustness against extrinsic noise $\sigma_{ex}^2$.

In the linear model, the robustness function $\delta_{q}(\sigma_{in}^2)$ can be derived analytically as:

\begin{align}
	\delta_{q}(\sigma_{in}^2)&=\frac{1}{\left(1+\sigma_{in}^{-2}\right)^{q}-1}-\sigma_{in}^2
	\label{LRM-the robustness function}
\end{align}
(see Appendix \ref{appendix-LRM-robustness-1}). The robustness function $\delta_{q}(\sigma_{in}^2)$ increases with intrinsic noise $\sigma_{in}^2$ for any $q$ (Fig. \ref{linear model}(d)). Therefore, INIR is clearly demonstrated. 

Moreover, the intrinsic noise dependency of the robustness function $\delta_{q}(\sigma_{in}^2)$ changes at $\sigma_{in}^2=1$ (Fig. \ref{linear model}(d)). The intrinsic noise dependency of the robustness function $\delta_{q}(\sigma_{in}^2)$ in $\sigma_{in}^2\gg1$ is stronger than that in $\sigma_{in}^2\ll1$. Indeed, Eq. (\ref{LRM-the robustness function}) can be approximated as follows:

\begin{align}
	\delta_{q}(\sigma_{in}^2)
	\propto\begin{cases}
		\sigma_{in}^{2q} & (\sigma_{in}^2\ll1)\\
		\sigma_{in}^2 & (\sigma_{in}^2\gg1)
	\end{cases}
	\label{LRM-the robustness function-approx}
\end{align}
(see Appendix \ref{appendix-LRM-robustness-2}). Considering $q\in(0,1)$, Eq. (\ref{LRM-the robustness function-approx}) indicates that the intrinsic noise dependency of the robustness function $\delta_{q}(\sigma_{in}^2)$ in $\sigma_{in}^2\gg1$ is stronger than that in $\sigma_{in}^2\ll1$.

As the input $X$ has variance 1, the intrinsic noise dependency of the robustness function $\delta_{q}(\sigma_{in}^2)$ becomes stronger when the intrinsic noise $\sigma_{in}^2$ is much larger than the variance of the input $X$. Therefore, the input distribution plays an important role in INIR.

\section{Threshold model\label{section-Threshold model}}
\subsection{The case $\boldsymbol{\theta=0}$\label{subsection-theta0}}
In Sec. \ref{section-Linear model}, we showed that the linear model realizes INIR. Moreover, we revealed that INIR is stronger when the intrinsic noise is much larger than the variance of the input. 

However, the input--output relation in cellular information transmission is often nonlinear. In particular, some form of threshold response is often observed \cite{Ozbudak2004, Melen2005, Narula2012}. Therefore, in this subsection, we introduce a threshold model and examine whether it reproduces the same results as the linear model. 

The input $X$, extrinsic noise $Z_{ex}$, and intrinsic noise $Z_{in}$ are assumed to obey Gaussian distributions $N(0,1)$, $N(0,\sigma_{ex}^2)$, and $N(0,\sigma_{in}^2)$, respectively, as for the linear model. However, the output $Y$ is given by:

\begin{align}
	Y&=\begin{cases}
		1 & (X+Z_{ex}+Z_{in}\geq\theta)\\
		0 & (X+Z_{ex}+Z_{in}<\theta)
	\end{cases}
	\label{Threshold response}
\end{align}
Therefore, the output $Y$ is 1 when $X+Z_{ex}+Z_{in}$ is greater than the threshold $\theta$, whereas the output $Y$ is 0 when $X+Z_{ex}+Z_{in}$ is less than $\theta$. In this paper, we call this the ``threshold model.'' 

In the threshold model, the input $X$ is a continuous random variable, but the output $Y$ is a discrete random variable. Therefore, the mutual information $I(X;Y)$ in the threshold model is given by \cite{Cover}:

\begin{align}
	I(X;Y)&=\sum_{Y}\int_{X}p(x,y)\log_{2}\frac{p(x,y)}{p(x)p(y)}dx
\end{align}

\begin{figure}[btp]
\begin{center}
	(a)\hspace{-3.0mm}
	\begin{minipage}[t][][b]{40mm}
		\includegraphics[width=40mm,height=35mm]{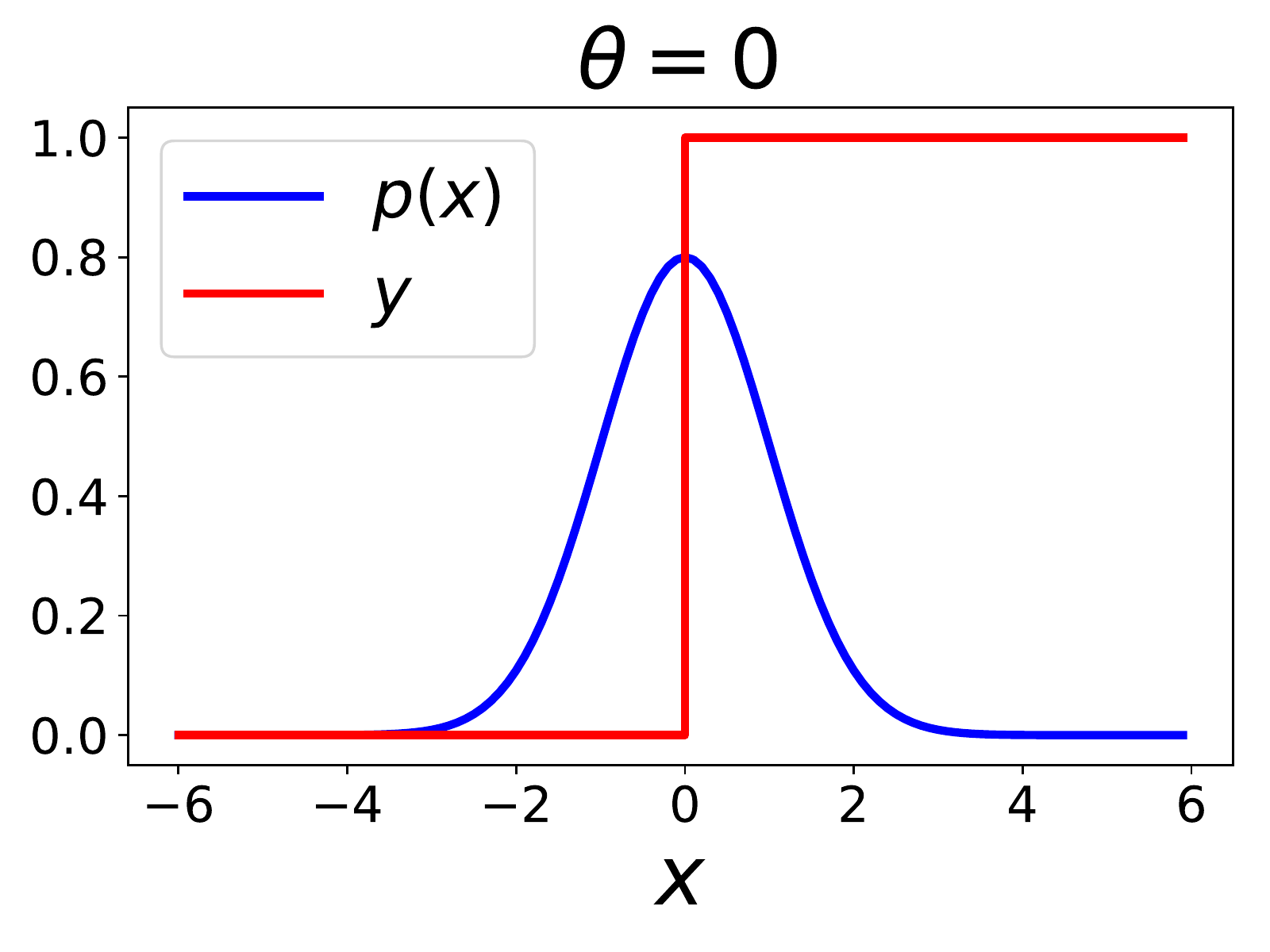}
	\end{minipage}
	(b)\hspace{-3.0mm}
	\begin{minipage}[t][][b]{40mm}
		\includegraphics[width=40mm,height=35mm]{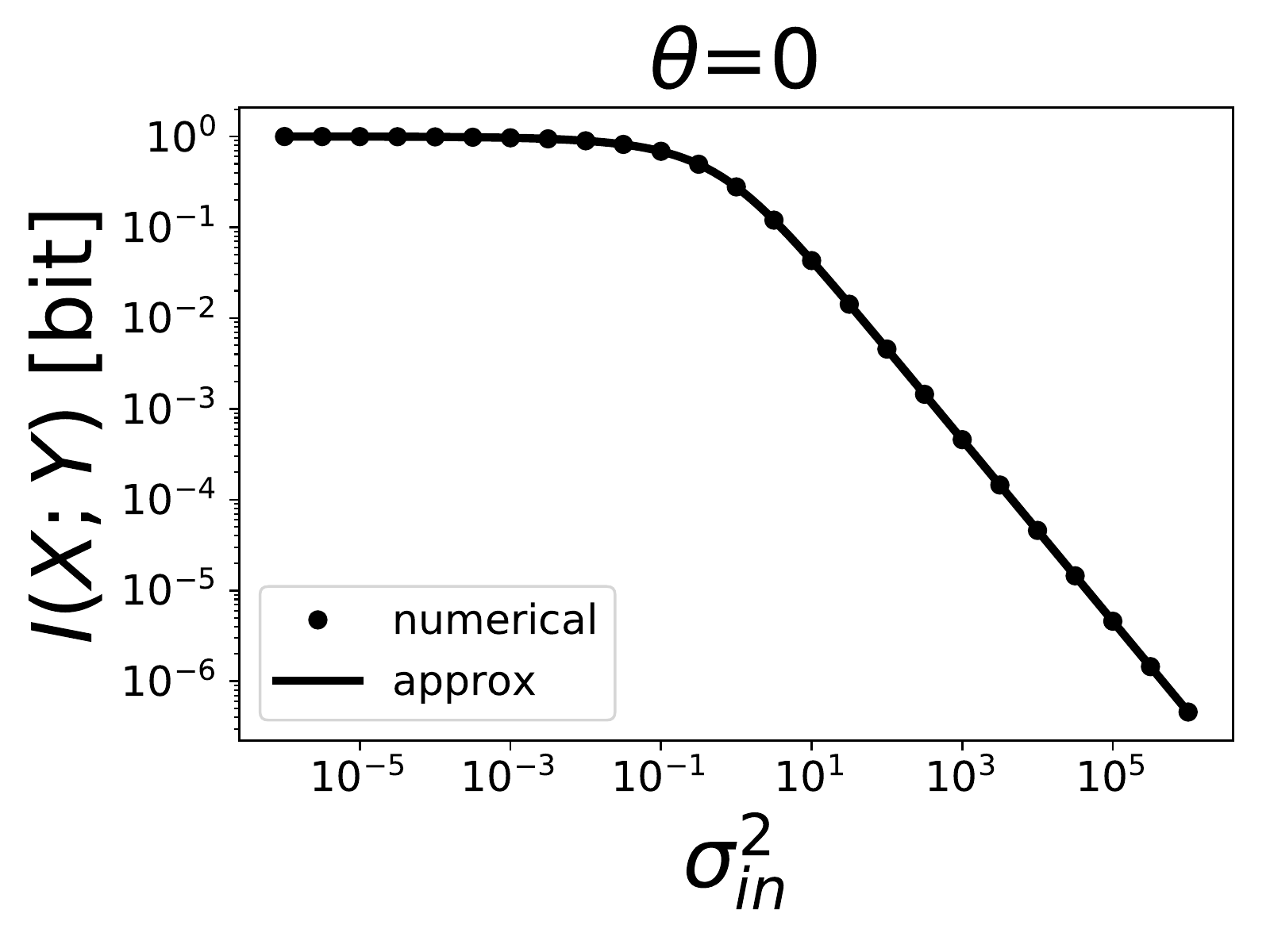}
	\end{minipage}\\
	(c)\hspace{-3.0mm}
	\begin{minipage}[t][][b]{40mm}
		\includegraphics[width=40mm,height=35mm]{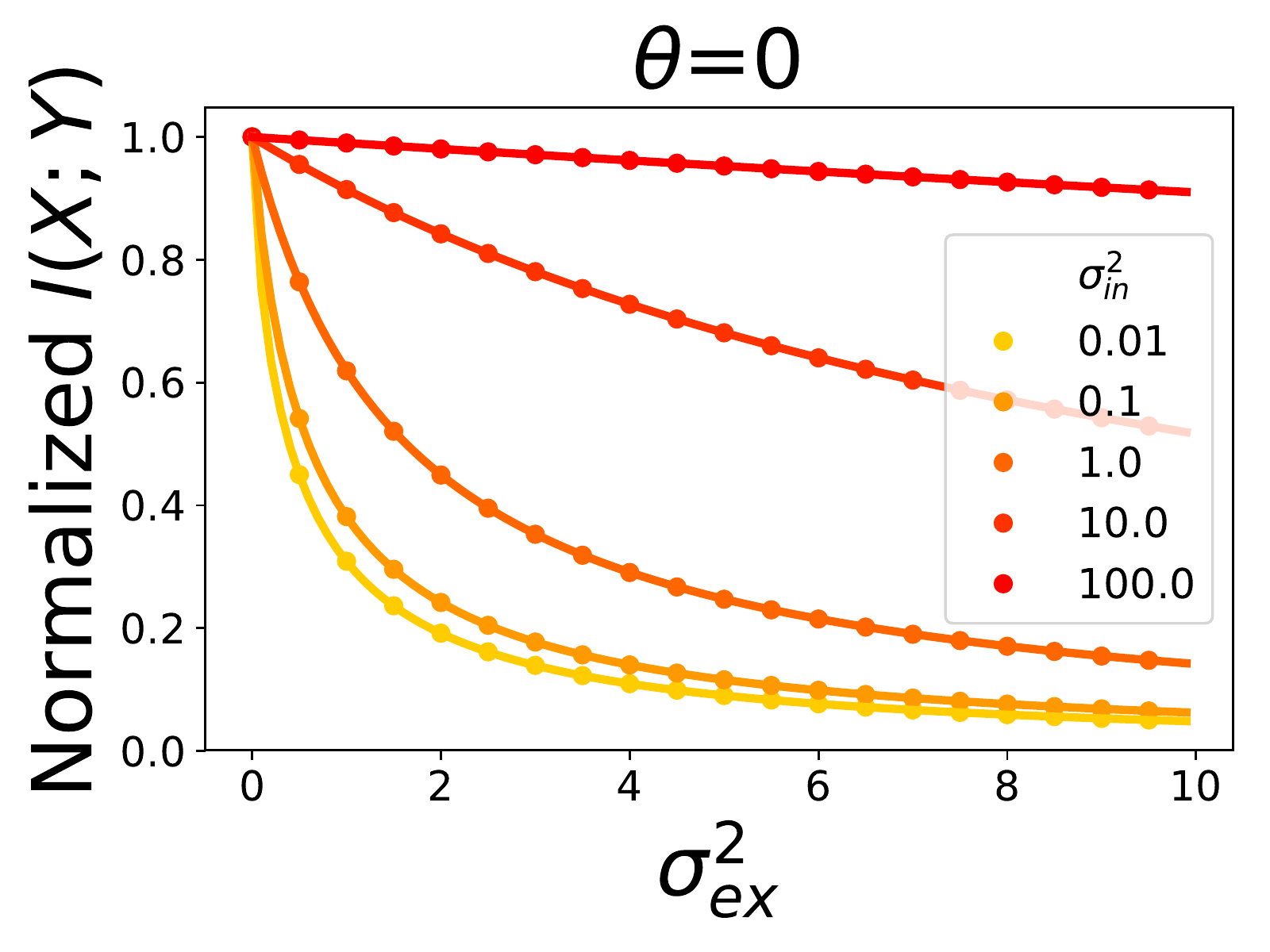}
	\end{minipage}
	(d)\hspace{-3.0mm}
	\begin{minipage}[t][][b]{40mm}
		\includegraphics[width=40mm,height=35mm]{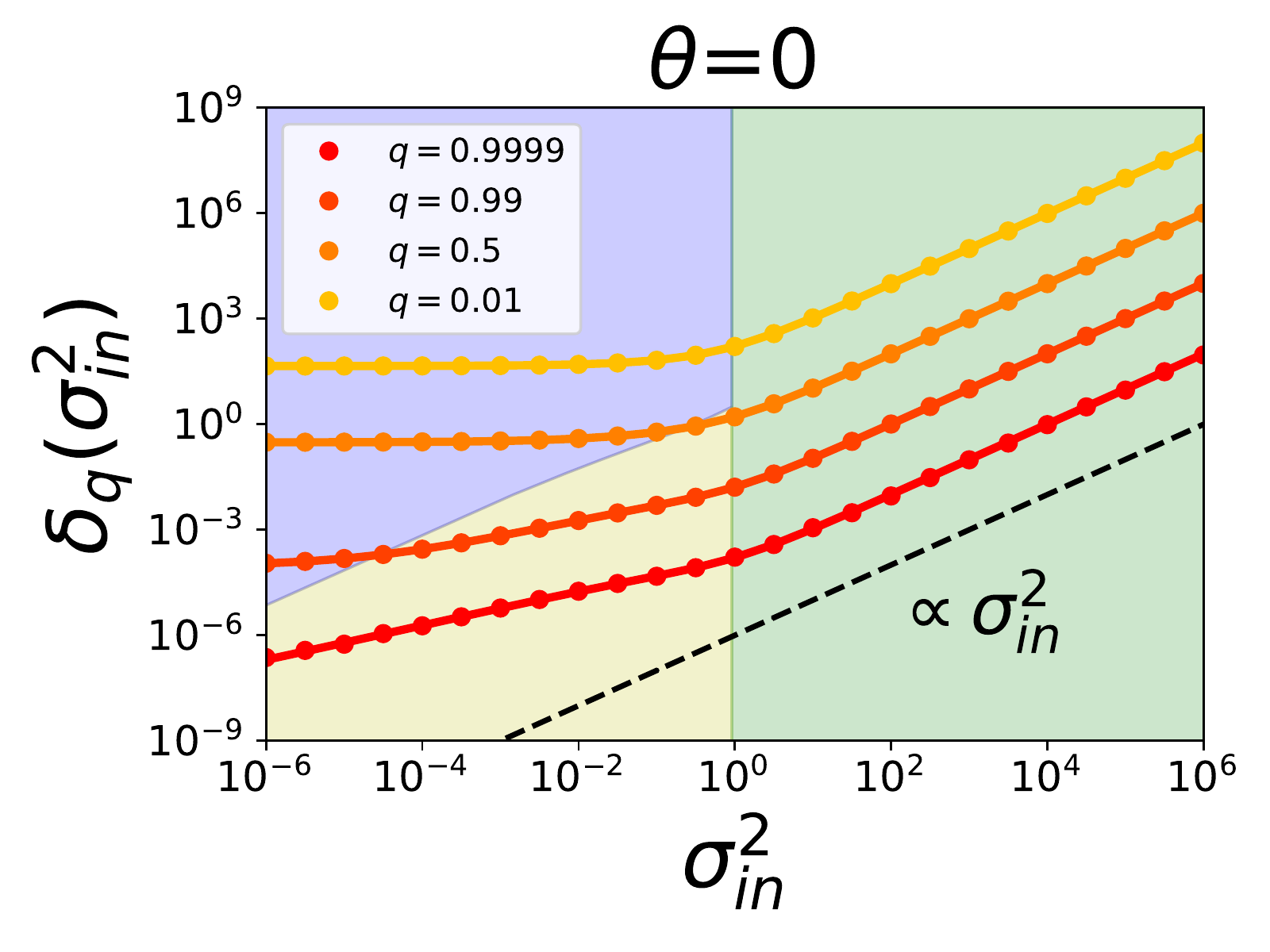}
	\end{minipage}
	 \caption{
	 Threshold model ($\theta=0$). 
	 (a) 
	 When $\theta=0$, the threshold $\theta$ is equal to the mean of the input $X$. 
	 (b)--(d)
	 Numerical solutions (dots) and approximate solutions (line).
	 (b) 
	 Mutual information $I(X;Y)$ against the intrinsic noise $\sigma_{in}^2$ when $\sigma_{ex}^2=0$. 
	 (c) 
	 Mutual information $I(X;Y)$ against extrinsic noise $\sigma_{ex}^2$. The mutual information $I(X;Y)$ is normalized at $\sigma_{ex}^2=0$.
	 (d)
	 The intrinsic noise dependency of the robustness function $\delta_{q}(\sigma_{in}^2)$. $\delta_{q}(\sigma_{in}^2)\propto\sigma_{in}^2$ when $\sigma_{in}^2>2/(\pi\ln2)$ (green), $\delta_{q}(\sigma_{in}^2)\propto\sigma_{in}$ when $((1-q)^2/(4q^2))(2/(\pi\ln2))<\sigma_{in}^2<2/(\pi\ln2)$ (yellow), and $\delta_{q}(\sigma_{in}^2)\propto{\rm const}$ when $\sigma_{in}^2<\min\{((1-q)^2/(4q^2))(2/(\pi\ln2)),2/(\pi\ln2)\}$ (blue). 
	 }
	\label{threshold model}
\end{center}
\end{figure}

For simplicity, in this subsection, we discuss the case of $\theta=0$ (Fig. \ref{threshold model}(a)). As input $X$ has a mean of 0, $\theta=0$ corresponds to the case where the mean of the input $X$ is equal to the threshold $\theta$. In Sec. \ref{subsection-SR}, we will discuss the case of $\theta\neq0$.

Unlike the linear model, the mutual information $I(X;Y)$ in the threshold model ($\theta=0$) cannot be derived analytically. However, it can be approximated as:

\begin{align}
	I(X;Y)&\approx1-\frac{1}{\sqrt{\frac{2}{(\pi\ln2)}\left(\sigma_{ex}^2+\sigma_{in}^2\right)^{-1}+1}}
	\label{TRM-theta0-MI}
\end{align}
(see Appendix \ref{appendix-TRM-MI-1}). This approximate solution matches the numerical solution closely (Fig. \ref{threshold model}(b), (c)).

The mutual information $I(X;Y)$ in the threshold model ($\theta=0$) decreases with the intrinsic noise $\sigma_{in}^2$ at $\sigma_{ex}^2=0$ (Fig. \ref{threshold model}(b)). However, the decrease in mutual information $I(X;Y)$ with respect to extrinsic noise $\sigma_{ex}^2$ occurs at a slower rate as the intrinsic noise $\sigma_{in}^2$ increases (Fig. \ref{threshold model}(c)). In other words, robustness against the extrinsic noise increases as the intrinsic noise becomes higher. Therefore, INIR is also realized in the threshold model ($\theta=0$). 

To investigate the robustness against extrinsic noise in more detail, we examine the robustness function $\delta_{q}(\sigma_{in}^2)$. As explained in Sec. \ref{section-Linear model}, the robustness function $\delta_{q}(\sigma_{in}^2)$ represents the extrinsic noise $\sigma_{ex}^2$ when the normalized mutual information $I(X;Y)/I(X;Y)|_{\sigma_{ex}^2=0}$ decreases to $q$ for $q\in(0,1)$. Large values of $\delta_{q}(\sigma_{in}^2)$ indicate high robustness against extrinsic noise.

The robustness function $\delta_{q}(\sigma_{in}^2)$ in the threshold model ($\theta=0$) cannot be derived analytically because we do not have an exact expression for $I(X;Y)$. However, $\delta_{q}(\sigma_{in}^2)$ can be derived approximately using the approximate solution of $I(X;Y)$ (Eq. (\ref{TRM-theta0-MI})). Thus:

\begin{align}
	\delta_{q}(\sigma_{in}^2)
	&\approx\frac{\left\{1-qI_{0}\right\}^2}{\left\{2-qI_{0}\right\}I_{0}}\frac{2}{(\pi\ln2)}-\sigma_{in}^2
	\label{TRM-the robustness function}
\end{align}
where 

\begin{align}
	I_{0}=1-\frac{1}{\sqrt{\frac{2}{(\pi\ln2)}\sigma_{in}^{-2}+1}}
	\label{TRM-the robustness function-I0}
\end{align}
(see Appendix \ref{appendix-TRM-robustness-1}). The robustness function $\delta_{q}(\sigma_{in}^2)$ increases with the intrinsic noise $\sigma_{in}^2$ for any $q$ (Fig. \ref{threshold model}(d)). Therefore, INIR is clearly demonstrated. 

Moreover, the intrinsic noise dependency of the robustness function $\delta_{q}(\sigma_{in}^2)$ changes at $\sigma_{in}^2=2/(\pi\ln2)$ (Fig. \ref{threshold model}(d)). The intrinsic noise dependency of the robustness function $\delta_{q}(\sigma_{in}^2)$ in $\sigma_{in}^2\gg2/(\pi\ln2)$ is stronger than that in $\sigma_{in}^2\ll2/(\pi\ln2)$. Furthermore, when $\sigma_{in}^2$ or $q$ is sufficiently small, the robustness function $\delta_{q}(\sigma_{in}^2)$ does not change with respect to the intrinsic noise $\sigma_{in}^2$. Indeed, Eq. (\ref{TRM-the robustness function}) can be approximated as follows:

\begin{align}
	\delta_{q}(\sigma_{in}^2)
	&\propto\begin{cases}
		\sigma_{in}^{2} & \left(\sigma_{in}^2\gg\frac{2}{\pi\ln2}\right)\\
		\sigma_{in} & \left(\frac{(1-q)^2}{4q^2}\frac{2}{\pi\ln2}\ll\sigma_{in}^2\ll\frac{2}{\pi\ln2}\right)\\
		{\rm const} & \left(\sigma_{in}^2\ll\min\left(\frac{(1-q)^2}{4q^2}\frac{2}{\pi\ln2},\frac{2}{\pi\ln2}\right)\right)
	\end{cases}
	\label{TRM-the robustness function-approx}
\end{align}
(see Appendix \ref{appendix-TRM-robustness-2}). Therefore, Eq. (\ref{TRM-the robustness function-approx}) indicates that the intrinsic noise dependency of the robustness function $\delta_{q}(\sigma_{in}^2)$ in $\sigma_{in}^2\gg2/(\pi\ln2)$ is stronger than that in $\sigma_{in}^2\ll2/(\pi\ln2)$. As the variance of the input $X$ is 1 and $2/(\pi\ln2)$ is close to 1, when the intrinsic noise is much larger than the variance of the input, the intrinsic noise dependency of the robustness against the extrinsic noise is particularly strong in the threshold model.

Furthermore, Eq. (\ref{TRM-the robustness function-approx}) suggests that the robustness function $\delta_{q}(\sigma_{in}^2)$ does not change when $\sigma_{in}^2\ll\min\{((1-q)^2/(4q^2))(2/(\pi\ln2)),2/(\pi\ln2)\}$. In particular, when $q\leq1/3$, the region in which $\delta_{q}(\sigma_{in}^2)\propto\sigma_{in}$ disappears, and $\delta_{q}(\sigma_{in}^2)$ remains constant when $\sigma_{in}^2\ll2/(\pi\ln2)$.

\subsection{The case $\boldsymbol{\theta\neq0}$\label{subsection-SR}}
\begin{figure*}[btp]
\begin{center}
	(a)\hspace{-3.0mm}
	\begin{minipage}[t][][b]{40mm}
		\includegraphics[width=40mm,height=35mm]{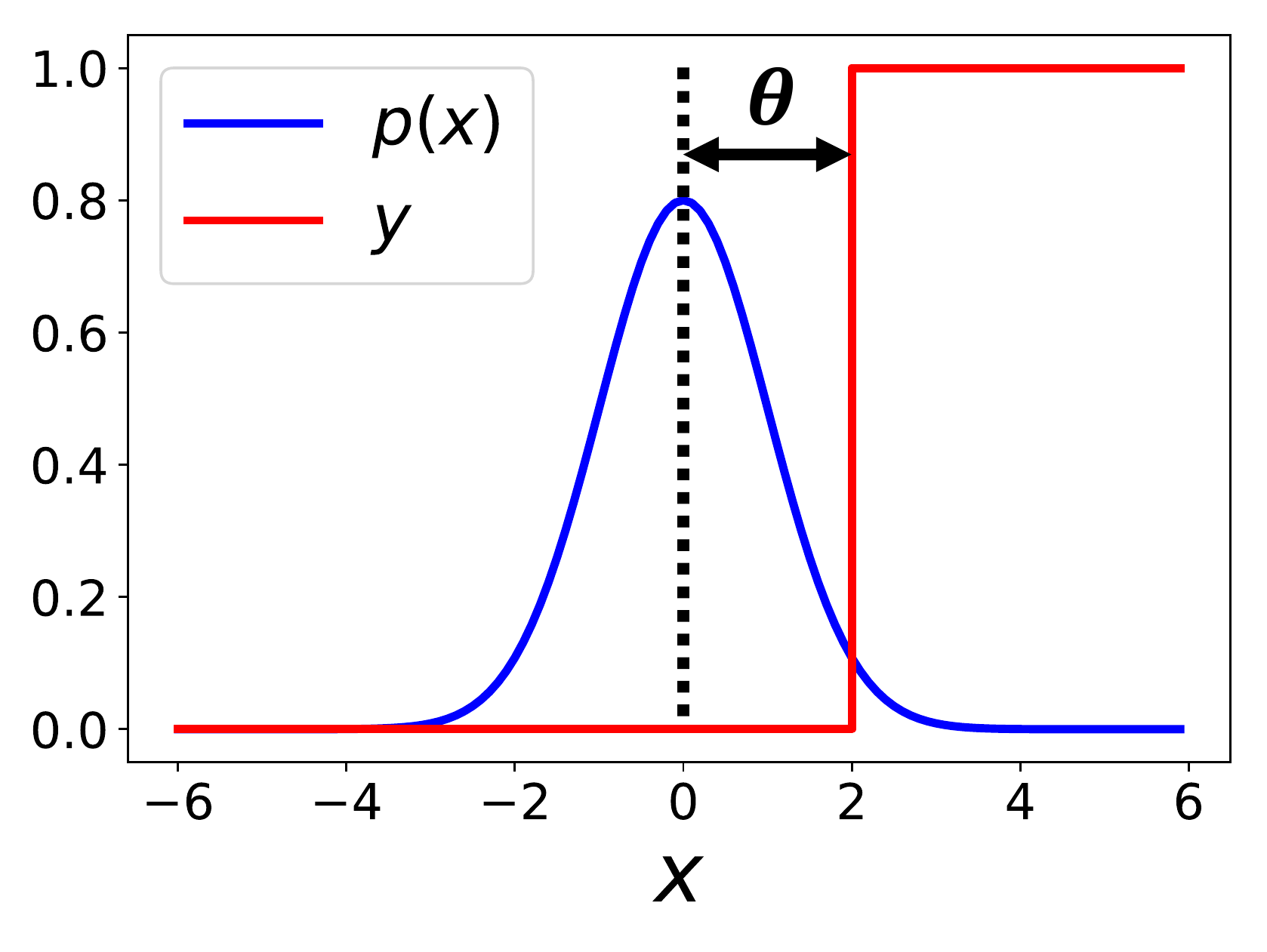}
	\end{minipage}
	(b)\hspace{-3.0mm}
	\begin{minipage}[t][][b]{50mm}
		\includegraphics[width=50mm,height=35mm]{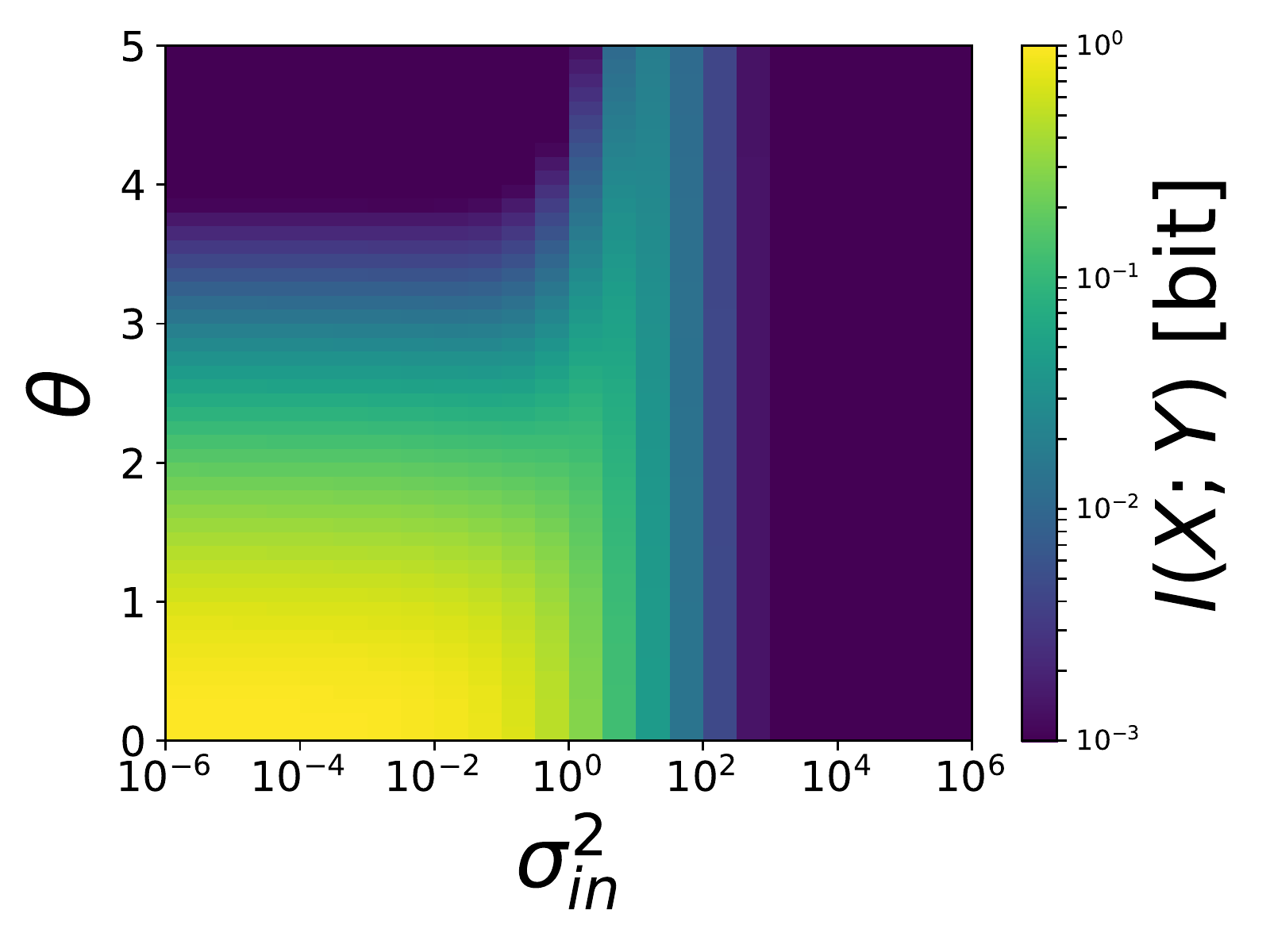}
	\end{minipage}
	(c)\hspace{-3.0mm}
	\begin{minipage}[t][][b]{40mm}
		\includegraphics[width=40mm,height=35mm]{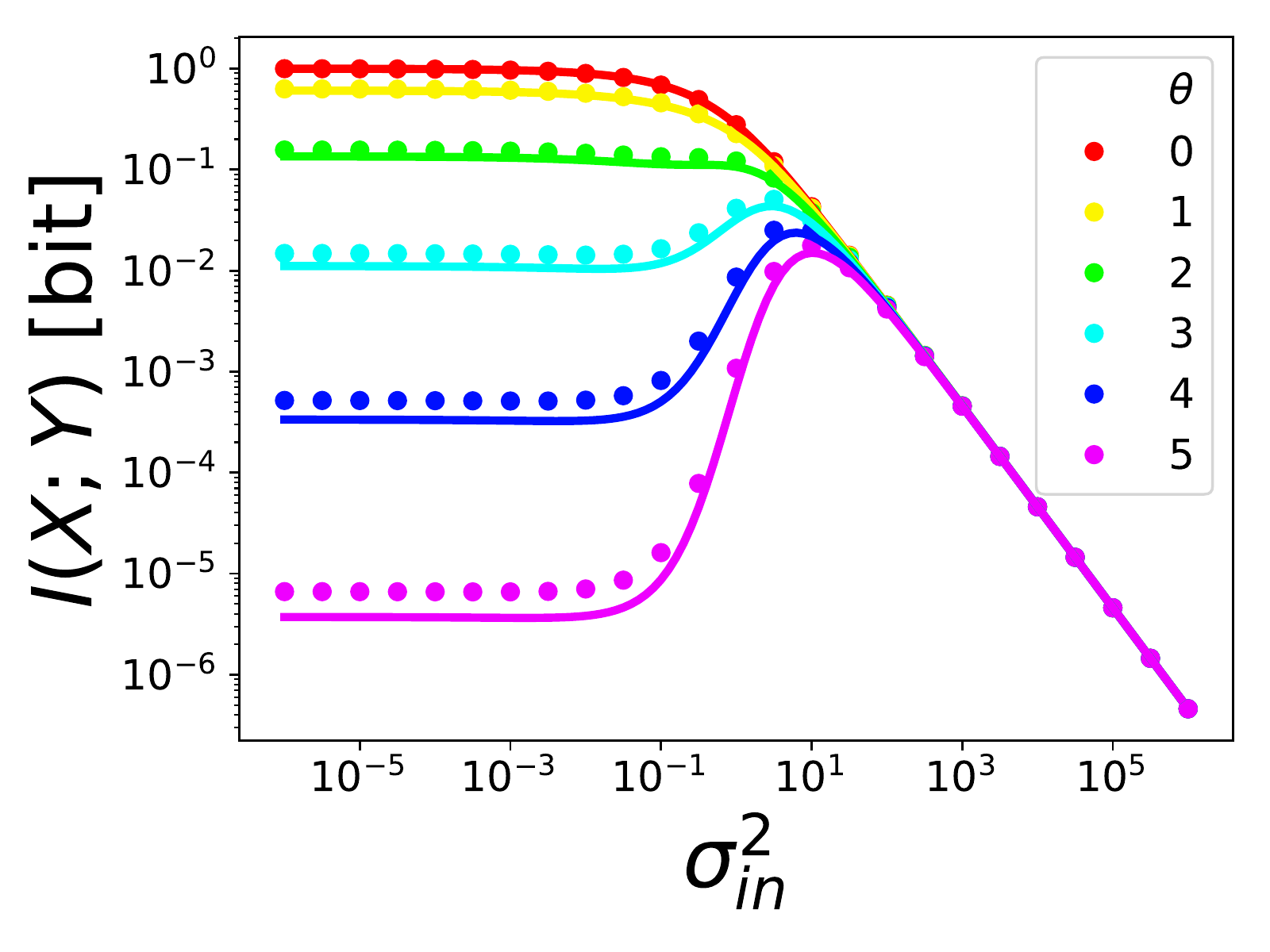}
	\end{minipage}\\
	(d)\hspace{-3.0mm}
	\begin{minipage}[t][][b]{40mm}
		\includegraphics[width=40mm,height=35mm]{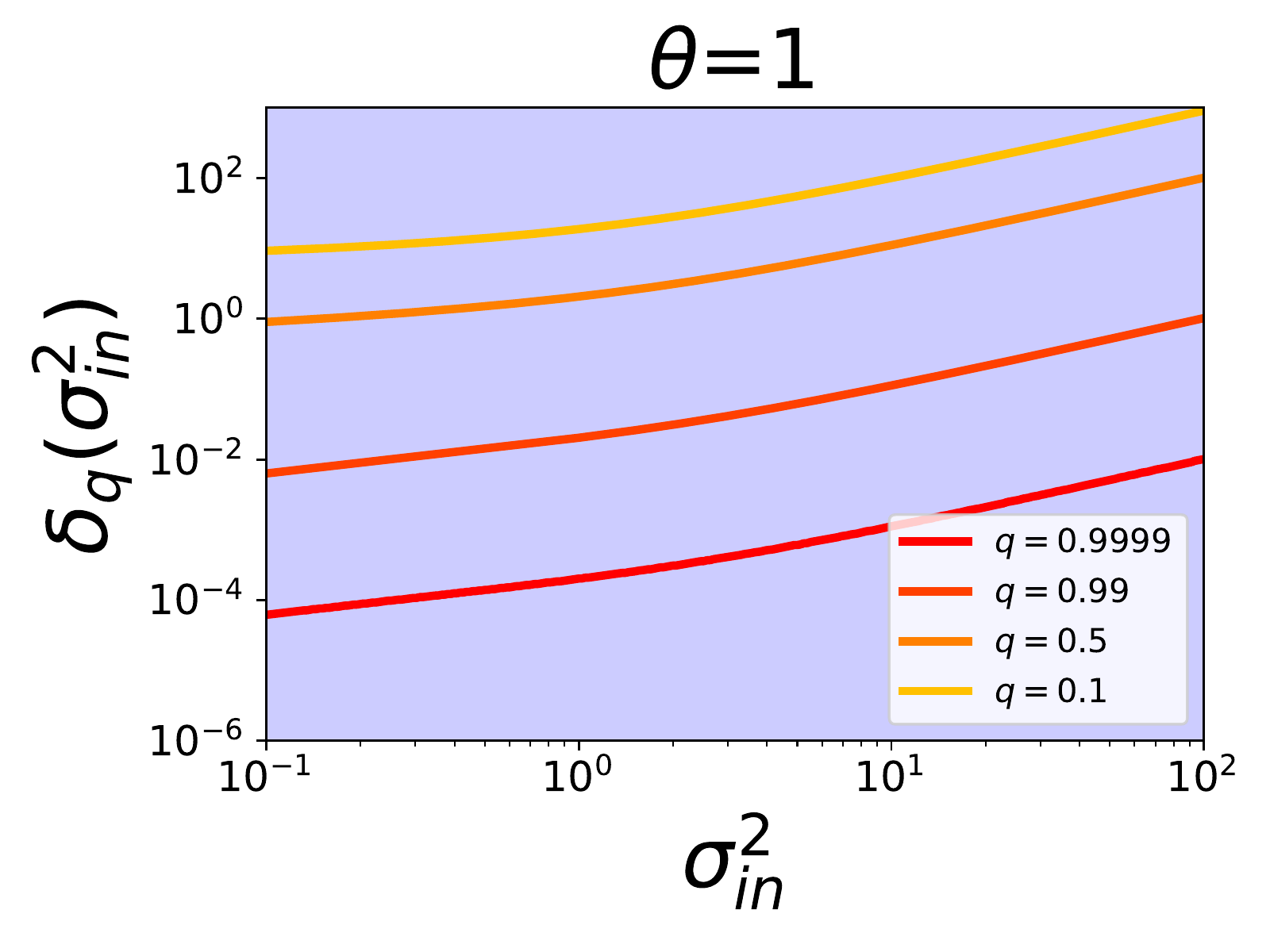}
	\end{minipage}
	(e)\hspace{-3.0mm}
	\begin{minipage}[t][][b]{40mm}
		\includegraphics[width=40mm,height=35mm]{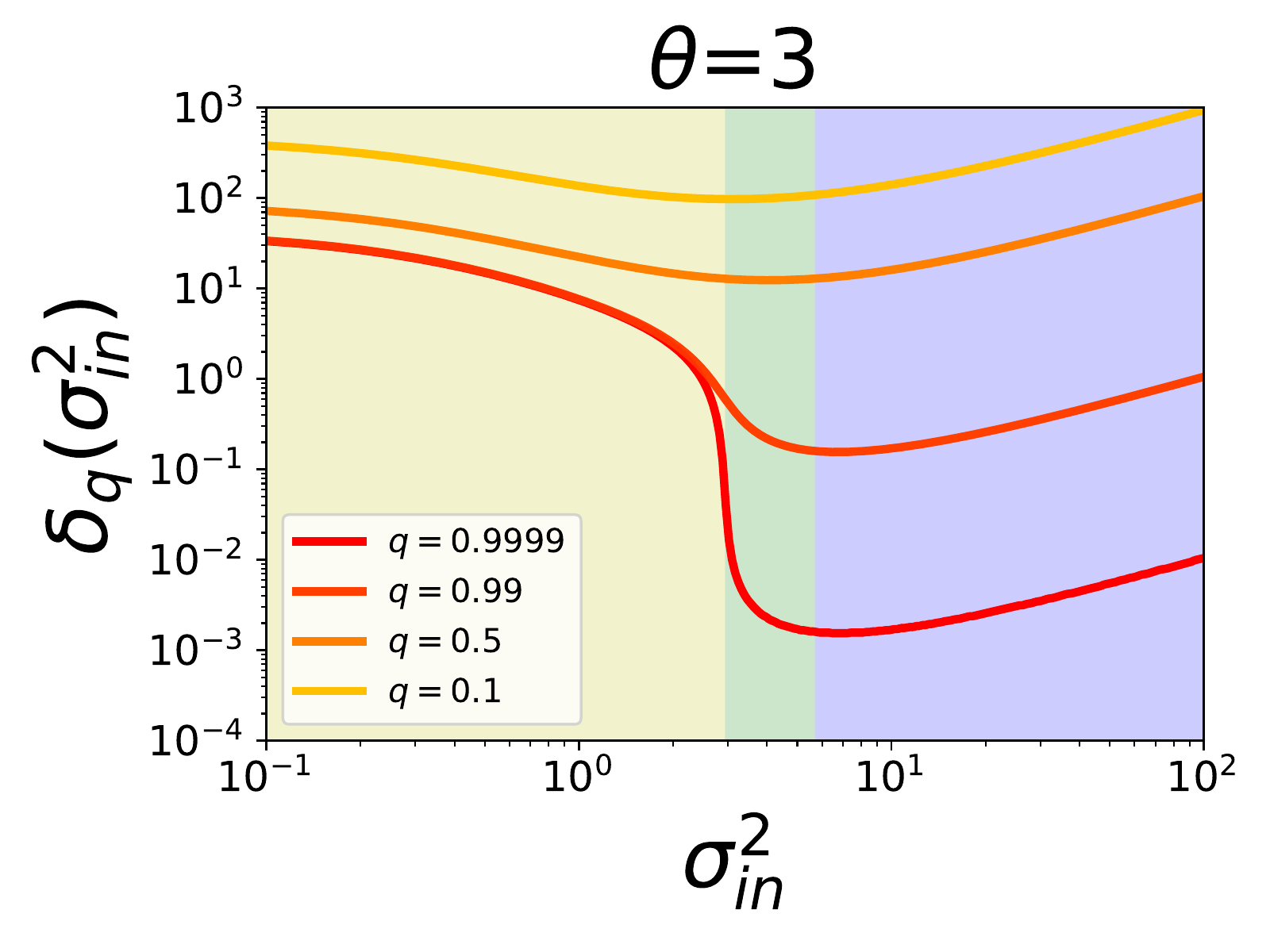}
	\end{minipage}
	(f)\hspace{-3.0mm}
	\begin{minipage}[t][][b]{40mm}
		\includegraphics[width=40mm,height=35mm]{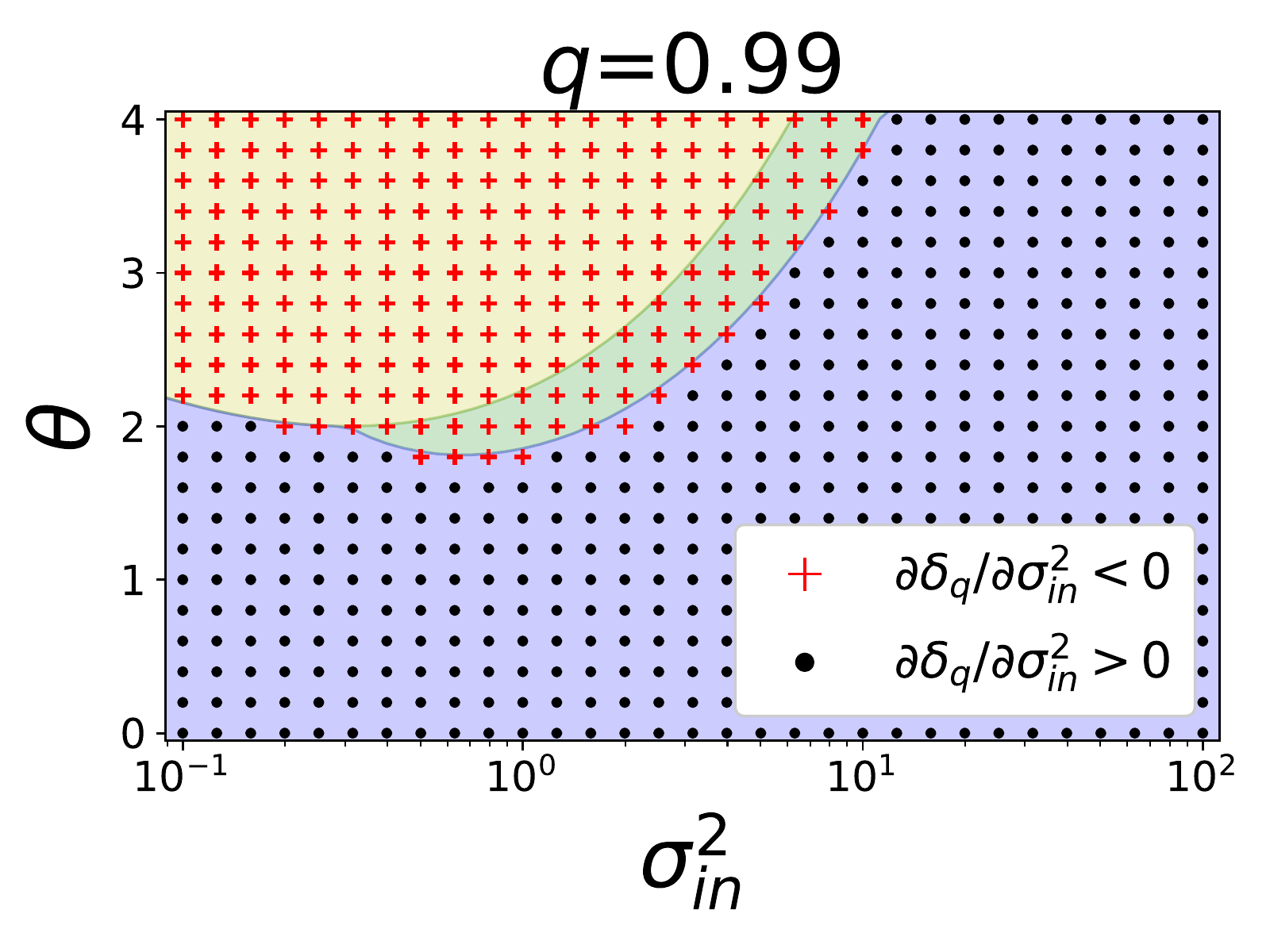}
	\end{minipage}
	(g)\hspace{-3.0mm}
	\begin{minipage}[t][][b]{40mm}
		\includegraphics[width=40mm,height=35mm]{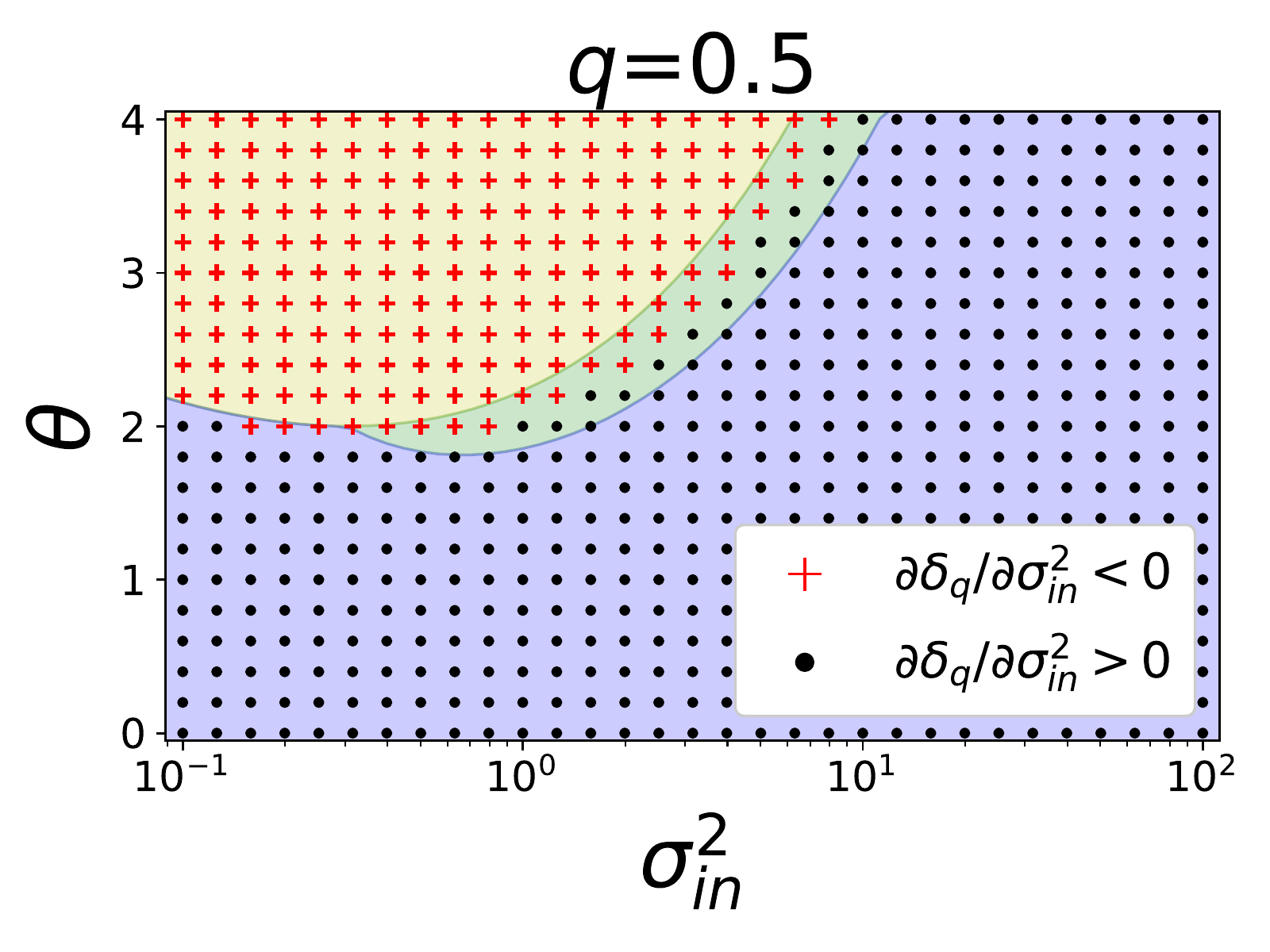}
	\end{minipage}
	\caption{
	 Threshold model ($\theta\neq0$). 
	 (a) When $\theta>0$, the threshold $\theta$ is larger than the mean of the input $X$.
	 (b), (c) Mutual information $I(X;Y)$ with respect to intrinsic noise $\sigma_{in}^2$ when $\sigma_{ex}^2=0$. (b) Numerical solutions. (c) Approximate solutions (lines). Numerical solutions (dots).
	 (d), (e) $\sigma_{in}^2$-dependency of the robustness function $\delta_{q}(\sigma_{in}^2)$.
	 (f), (g) $\sigma_{in}^2$, $\theta$-dependency of $\partial\delta_{q}/\partial\sigma_{in}^2$. 
 	 (d)--(g) $\partial I(X;Y)/\partial \sigma_{in}^2>0$ (yellow), $\partial I(X;Y)/\partial \sigma_{in}^2<0$ and $\partial^2 I(X;Y)/(\partial \sigma_{in}^2)^2<0$ (green), $\partial I(X;Y)/\partial \sigma_{in}^2<0$ and $\partial^2 I(X;Y)/(\partial \sigma_{in}^2)^2>0$ (blue).
	 }
	\label{threshold model SR}
\end{center}
\end{figure*}

In Sec. \ref{subsection-theta0}, we discussed the threshold model in the case $\theta=0$. In this subsection, we consider the case $\theta\neq0$ (Fig. \ref{threshold model SR}(a)). The mutual information $I(X;Y)$ is an even function of $\theta$. Therefore, we can consider the case $\theta\geq0$ without loss of generality.

The mutual information $I(X;Y)$ in the threshold model cannot be derived analytically, but the following approximate expression can be obtained:

\begin{align}
	I(X;Y)\approx&\left(1-\frac{1}{\sqrt{\frac{2}{(\pi\ln2)}\left(\sigma_{ex}^2+\sigma_{in}^2\right)^{-1}+1}}\right)\nonumber\\
	&\times\exp\left(-\frac{\theta^2}{(\pi\ln2)\left(\frac{2}{(\pi\ln2)}+\sigma_{ex}^2+\sigma_{in}^{2}\right)}\right)
	\label{TRM-SR-MI}
\end{align}
(see Appendix \ref{appendix-TRM-SR-MI-1}). This approximate solution matches the numerical solution well (Fig. \ref{threshold model SR}(c)).

When $\theta=0$, that is, the threshold $\theta$ is equal to the mean of the input $X$, the mutual information $I(X;Y)$ decreases monotonically with respect to the intrinsic noise $\sigma_{in}^2$ (Fig. \ref{threshold model SR}(c)(red)). When $\theta$ is much larger than 0, that is, the threshold $\theta$ is much larger than the mean of the input $X$, the mutual information $I(X;Y)$ is maximized with a moderate value of the intrinsic noise $\sigma_{in}^2$ (Fig. \ref{threshold model SR}(c)). This phenomenon is called ``stochastic resonance'' (SR) \cite{Stocks2000, Stocks2001}.

Why does SR appear in the threshold model for large $\theta$? This is the same as asking why the mutual information $I(X;Y)$ is maximized with a moderate intrinsic noise $\sigma_{in}^2$ for large $\theta$. This mechanism can be explained as follows. When $\theta$ is large, that is, the threshold $\theta$ is much larger than the mean of the input $X$, the input $X$ barely exceeds the threshold $\theta$. Therefore, little of the input information is transmitted to the output in the absence of intrinsic noise $\sigma_{in}^2$. However, when the intrinsic noise $\sigma_{in}^2$ is added to the input $X$, the width of the input distribution increases, and the number of times the input $X$ exceeds the threshold $\theta$ increases. Therefore, the transmitted input information increases. Nonetheless, when the intrinsic noise $\sigma_{in}^2$ is too large, the input information is hidden by the large noise, and in this case, the transmitted input information decreases. Through this mechanism, the mutual information $I(X;Y)$ can be maximized with a moderate intrinsic noise $\sigma_{in}^2$ at large $\theta$.

Stocks has already reported that SR appears in the same threshold model \cite{Stocks2000, Stocks2001}. However, the condition under which SR appears has only been illustrated through numerical calculations. We can derive the approximate condition under which SR appears: $\theta>2$ (see Appendix \ref{appendix-TRM-SR-MI-2}). As the length of $X$ is scaled by its variance, SR appears when the threshold $\theta$ is more than twice the variance of the input $X$, which is consistent with the numerical results from a previous study \cite{Munakata2005}. 

To investigate the robustness of the threshold model against extrinsic noise at $\theta>0$, we examine the robustness function $\delta_{q}(\sigma_{in}^2)$ in the threshold model at $\theta=1$ (Fig. \ref{threshold model SR}(d)) and at $\theta=3$ (Fig. \ref{threshold model SR}(e)). As mentioned in the previous sections, when the robustness function $\delta_{q}(\sigma_{in}^2)$ is large, the robustness against extrinsic noise $\sigma_{ex}^2$ is high. When $\theta>0$, we cannot derive the robustness function $\delta_{q}(\sigma_{in}^2)$ analytically. Therefore, we derive $\delta_{q}(\sigma_{in}^2)$ numerically (Fig. \ref{threshold model SR}(d), (e)). 

When SR does not appear ($\theta=1$), the robustness function $\delta_{q}(\sigma_{in}^2)$ increases monotonically with the intrinsic noise $\sigma_{in}^2$ for any $q$ (Fig. \ref{threshold model SR}(d)). However, when SR appears ($\theta=3$), the robustness function $\delta_{q}(\sigma_{in}^2)$ does not always increase with the intrinsic noise $\sigma_{in}^2$ (Fig. \ref{threshold model SR}(e)). Therefore, INIR is not always realized in the presence of SR. 

We further examine the region where INIR is not realized. Defining $\sigma_{in,\max}^2$ as the intrinsic noise $\sigma_{in}^2$ that maximizes the mutual information $I(X;Y)$, the robustness function $\delta_{q}(\sigma_{in}^2)$ decreases with the intrinsic noise when $\sigma_{in}^2<\sigma_{in,\max}^2$ (Fig. \ref{threshold model SR}(e)(yellow)). In other words, the robustness against extrinsic noise $\sigma_{ex}^2$ decreases with respect to the intrinsic noise $\sigma_{in}^2$ when the mutual information $I(X;Y)$ increases with respect to $\sigma_{in}^2$. This is trivial, because the mutual information $I(X;Y)$ first increases and then decreases with extrinsic noise $\sigma_{ex}^2$ when $\sigma_{in}^2<\sigma_{in,\max}^2$. Therefore, when the intrinsic noise $\sigma_{in}^2$ decreases in $\sigma_{in}^2<\sigma_{in,\max}^2$, the range of the extrinsic noise $\sigma_{ex}^2$ for which the mutual information $I(X;Y)$ increases will expand. Thus, the robustness against extrinsic noise $\sigma_{ex}^2$ increases as the intrinsic noise $\sigma_{in}^2$ decreases when $\sigma_{in}^2<\sigma_{in,\max}^2$.

Interestingly, even when $\sigma_{in}^2>\sigma_{in,\max}^2$, there is a region where the robustness function $\delta_{q}(\sigma_{in}^2)$ decreases with the intrinsic noise $\sigma_{in}^2$ (Fig. \ref{threshold model SR}(e)). In other words, the robustness against the extrinsic noise $\sigma_{ex}^2$ decreases with respect to the intrinsic noise $\sigma_{in}^2$ even when the mutual information $I(X;Y)$ decreases with respect to $\sigma_{in}^2$. Defining $\sigma_{in,c}^2$ as the intrinsic noise $\sigma_{in}^2$ at which the mutual information $I(X;Y)$ changes from convex upward to convex downward, the robustness function $\delta_{q}(\sigma_{in}^2)$ decreases with respect to the intrinsic noise $\sigma_{in}^2$ when $\sigma_{in,\max}^2<\sigma_{in}^2<\sigma_{in,c}^2$ and $q$ is close to 1 (Fig. \ref{threshold model SR}(e)(green)). In other words, the robustness against the extrinsic noise $\sigma_{ex}^2$ decreases as the intrinsic noise $\sigma_{in}^2$ increases when the mutual information $I(X;Y)$ is decreasing and convex upward with respect to $\sigma_{in}^2$.

We have approximately demonstrated that the robustness function $\delta_{q}(\sigma_{in}^2)$ decreases with respect to the intrinsic noise $\sigma_{in}^2$ when the mutual information $I(X;Y)$ is decreasing and convex upward against the intrinsic noise $\sigma_{in}^2$, and $q$ is close to 1.

In the threshold model, the $\sigma_{in}^2$-dependency of $I(X;Y)$ is the same as the $\sigma_{ex}^2$-dependency of $I(X;Y)$, and $I(X;Y)$ is decreasing against $\sigma_{in}^2$. Therefore, 

\begin{align}
	\frac{\partial I(X;Y)}{\partial \sigma_{ex}^2}=\frac{\partial I(X;Y)}{\partial \sigma_{in}^2}<0
\end{align}
is satisfied. When $\partial I(X;Y)/\partial \sigma_{ex}^2<0$, the robustness function $\delta_{q}(\sigma_{in}^2)$ becomes close to 0 as $q\to1$. Therefore, when $\partial I(X;Y)/\partial \sigma_{ex}^2<0$ and $q\sim1$,

\begin{align}
	&\left.I(X;Y)\right|_{\sigma_{ex}^2=\delta_{q}}\nonumber\\
	&=\left.I(X;Y)\right|_{\sigma_{ex}^2=0}+\left.\frac{\partial I(X;Y)}{\partial\sigma_{ex}^2}\right|_{\sigma_{ex}^2=0}\delta_{q}+O\left(\delta_{q}^2\right)
	\label{Taylor expansion of MI}
\end{align}
is satisfied. Substituting Eq. (\ref{Taylor expansion of MI}) into Eq. (\ref{the definition of the robustness function}), we obtain

\begin{align}
	\delta_{q}(\sigma_{in}^2)&=-\frac{(1-q)}{\left.\frac{\partial\ln I(X;Y)}{\partial\sigma_{ex}^2}\right|_{\sigma_{ex}^2=0}}
	\label{the robustness function at q1}
\end{align}
$\partial\ln I(X;Y)/\partial\sigma_{ex}^2|_{\sigma_{ex}^2=0}$ corresponds to the slope of the normalized mutual information $I(X;Y)/I(X;Y)|_{\sigma_{ex}^2=0}$ against the extrinsic noise $\sigma_{ex}^2$ at $\sigma_{ex}^2=0$. When $\partial\ln I(X;Y)/\partial\sigma_{ex}^2|_{\sigma_{ex}^2=0}$ is large, the robustness function $\delta_{q}(\sigma_{in}^2)$ at $q\sim1$ is also large. 

Differentiating Eq. (\ref{the robustness function at q1}) with respect to the intrinsic noise $\sigma_{in}^2$, we have

\begin{align}
	\frac{\partial \delta_{q}(\sigma_{in}^2)}{\partial \sigma_{in}^2}
	&=\frac{(1-q)\frac{\partial}{\partial\sigma_{in}^2}\left(\left.\frac{\partial \ln I(X;Y)}{\partial\sigma_{ex}^2}\right|_{\sigma_{ex}^2=0}\right)}{\left(\left.\frac{\partial \ln I(X;Y)}{\partial\sigma_{ex}^2} \right|_{\sigma_{ex}^2=0}\right)^2}
	\label{the robustness function differentiated by the intrinsic noise}
\end{align}
In the threshold model, $I(X;Y)$ depends on the summation of $\sigma_{in}^2$ and $\sigma_{ex}^2$. Therefore, 

\begin{align}
	\frac{\partial}{\partial\sigma_{in}^2}\left(\left.\frac{\partial \ln I(X;Y)}{\partial\sigma_{ex}^2}\right|_{\sigma_{ex}^2=0}\right)
	=\left.\frac{\partial^2 \ln I(X;Y)}{\left(\partial\sigma_{in}^2\right)^2}\right|_{\sigma_{ex}^2=0}
	\label{the mutual information differentiated by the intrinsic noise twice}
\end{align}
is satisfied. From Eqs. (\ref{the robustness function differentiated by the intrinsic noise}) and (\ref{the mutual information differentiated by the intrinsic noise twice}), and for $q<1$, 

\begin{align}
	{\rm sgn}\left(\frac{\partial \delta_{q}(\sigma_{in}^2)}{\partial \sigma_{in}^2}\right)
	&={\rm sgn}\left(\left.\frac{\partial^2 \ln I(X;Y)}{\left(\partial\sigma_{in}^2\right)^2}\right|_{\sigma_{ex}^2=0}\right)
	\label{relation between delta robustness function and delta mutual information}
\end{align}
is satisfied, where ${\rm sgn}(\cdot)$ is the sign function. 

Here, from simple calculations, 

\begin{align}
	\frac{\partial^2 I(X;Y)}{\left(\partial\sigma_{in}^2\right)^2}<0\Rightarrow\frac{\partial^2 \ln I(X;Y)}{\left(\partial\sigma_{in}^2\right)^2}<0
\end{align}
is satisfied. Therefore, from Eq. (\ref{relation between delta robustness function and delta mutual information}), 

\begin{align}
	\frac{\partial^2 I(X;Y)}{\left(\partial\sigma_{in}^2\right)^2}<0\Rightarrow\frac{\partial \delta_{q}(\sigma_{in}^2)}{\partial \sigma_{in}^2}<0
	\label{equation-green-region}
\end{align}
is satisfied. Therefore, when $I(X;Y)$ is decreasing and convex upward with respect to $\sigma_{in}^2$, the robustness function $\delta_{q}(\sigma_{in}^2)$ at $q\sim1$ decreases with $\sigma_{in}^2$. Moreover, 

\begin{align}
	\frac{\partial^2 \ln I(X;Y)}{\left(\partial\sigma_{in}^2\right)^2}>0\Rightarrow\frac{\partial^2 I(X;Y)}{\left(\partial\sigma_{in}^2\right)^2}>0
\end{align}
is satisfied. Therefore, from Eq. (\ref{relation between delta robustness function and delta mutual information}), 

\begin{align}
	\frac{\partial \delta_{q}(\sigma_{in}^2)}{\partial \sigma_{in}^2}>0\Rightarrow\frac{\partial^2 I(X;Y)}{\left(\partial\sigma_{in}^2\right)^2}>0
	\label{equation-blue-region}
\end{align}
is also satisfied. Hence, when $I(X;Y)$ is decreasing against $\sigma_{in}^2$, and $\delta_{q}(\sigma_{in}^2)$ at $q\sim1$ is increasing against $\sigma_{in}^2$, $I(X;Y)$ is always convex downward with respect to $\sigma_{in}^2$. 

We now verify Eqs. (\ref{equation-green-region}) and (\ref{equation-blue-region}) numerically (Fig. \ref{threshold model SR}(f),(g)). When $q$ is close to 1, Eqs. (\ref{equation-green-region}) and (\ref{equation-blue-region}) are satisfied (Fig. \ref{threshold model SR}(f)). However,  when $q$ is far from 1, Eqs. (\ref{equation-green-region}) and (\ref{equation-blue-region}) are not satisfied (Fig. \ref{threshold model SR}(g)). These results are as expected. 

Therefore, it has been shown that the robustness function $\delta_{q}(\sigma_{in}^2)$ at $q\sim1$ decreases with the intrinsic noise $\sigma_{in}^2$ when $I(X;Y)$ is decreasing and convex upward against $\sigma_{in}^2$ (Fig. \ref{threshold model SR}(f)(green)). Thus, INIR is not always realized in the threshold model with SR, even when $I(X;Y)$ is decreasing against $\sigma_{in}^2$. However, this is not necessarily bad, and may indeed be an advantageous property. In the linear model and the threshold model without SR, reducing the intrinsic noise $\sigma_{in}^2$ increases the mutual information $I(X;Y)$, but decreases the robustness function $\delta_{q}(\sigma_{in}^2)$. Hence, there is a trade-off between the transmitted information and the robustness against extrinsic noise in the linear model and the threshold model without SR. In contrast, in the threshold model with SR, the mutual information $I(X;Y)$ and the robustness function $\delta_{q}(\sigma_{in}^2)$ at $q\sim1$ can be simultaneously increased by reducing the intrinsic noise $\sigma_{in}^2$ when $I(X;Y)$ is decreasing and convex upward with respect to $\sigma_{in}^2$ (Fig. \ref{threshold model SR}(f)(green)). Therefore, SR solves the trade-off problem between the transmitted information and the robustness against extrinsic noise.

\section{Mechanism\label{section-Mechanism}}
\begin{figure}[btp]
\begin{center}
	(a)\hspace{-3.0mm}
	\begin{minipage}[t][][b]{40mm}
		\includegraphics[width=40mm,height=35mm]{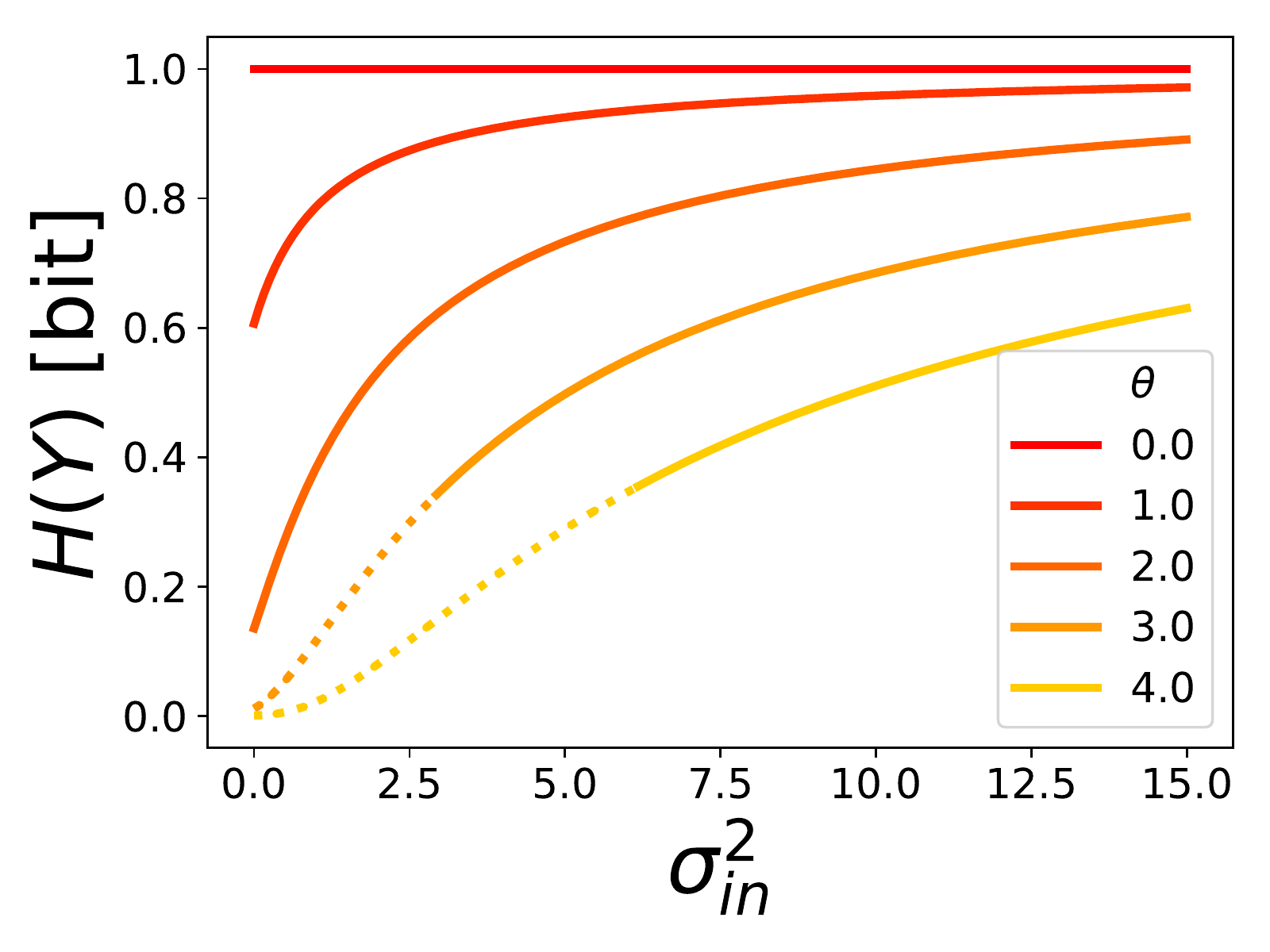}
	\end{minipage}
	(b)\hspace{-3.0mm}
	\begin{minipage}[t][][b]{40mm}
		\includegraphics[width=40mm,height=35mm]{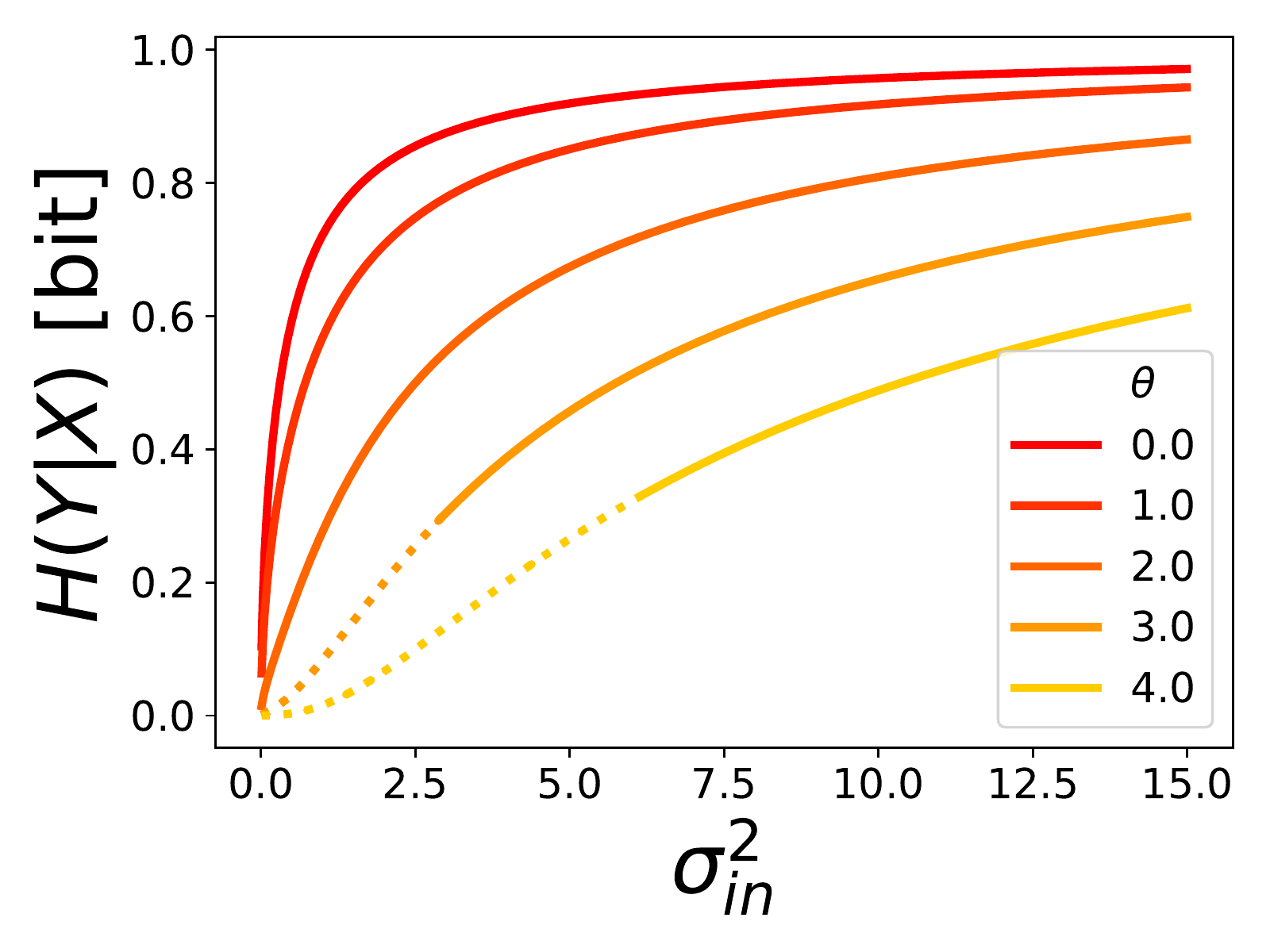}
	\end{minipage}
	 \caption{
	 Threshold model.  $\sigma_{in}^2$-dependencies of (a) $H(Y)$ and (b)  $H(Y|X)$. $\partial I(X;Y)/\partial \sigma_{in}^2<0$ (solid line), $\partial I(X;Y)/\partial \sigma_{in}^2>0$ (dashed line).
	 }
	\label{entropy}
\end{center}
\end{figure}

In Sec. \ref{subsection-SR}, we showed that when $\partial I(X;Y)/\partial\sigma_{in}^2<0$ and $q\sim1$, the sign of $\partial\delta_{q}(\sigma_{in}^2)/\partial\sigma_{in}^2$ strongly depends on the sign of $\partial^2 I(X;Y)/(\partial\sigma_{in}^2)^2$. The robustness against extrinsic noise increases with respect to the intrinsic noise when $\partial^2 I(X;Y)/(\partial\sigma_{in}^2)^2>0$ (Fig. \ref{threshold model SR}(f)(blue)), whereas it decreases when $\partial^2 I(X;Y)/(\partial\sigma_{in}^2)^2<0$ (Fig. \ref{threshold model SR}(f)(green)). In this subsection, we explain the relation between $\partial\delta_{q}(\sigma_{in}^2)/\partial\sigma_{in}^2$ and $\partial^2 I(X;Y)/(\partial\sigma_{in}^2)^2$ more intuitively by decomposing $I(X;Y)$.

The mutual information $I(X;Y)$ is decomposed as follows:

\begin{align}
	I(X;Y)&=H(Y)-H(Y|X)
	\label{the decomposition of the mutual information into entropies}
\end{align}
where $H(Y)$ is the entropy of $Y$ and $H(Y|X)$ is the entropy of $Y$ given by $X$, which is expressed as:

\begin{align}
	H(Y)&=-\sum_{Y}p(y)\log_{2}p(y)\\
	H(Y|X)&=-\int_{X}p(x)\sum_{Y}p(y|x)\log_{2}p(y|x)dx
\end{align}
where $X$ is a continuous random variable and $Y$ is a discrete random variable.

In the threshold model, when $\partial I(X;Y)/\partial\sigma_{in}^2<0$, both $H(Y)$ and $H(Y|X)$ are increasing and convex upward with respect to the intrinsic noise $\sigma_{in}^2$ (Fig. \ref{entropy}(a), (b)(solid line)), and thus satisfy

\begin{align}
	&\frac{\partial H(Y)}{\partial \sigma_{in}^2}\geq0,&\frac{\partial H(Y|X)}{\partial \sigma_{in}^2}\geq0
	\label{first derivative of H(Y) and H(Y|X)}\\
	&\frac{\partial^2 H(Y)}{\left(\partial \sigma_{in}^2\right)^2}\leq0,&\frac{\partial^2 H(Y|X)}{\left(\partial \sigma_{in}^2\right)^2}\leq0
	\label{second derivative of H(Y) and H(Y|X)}
\end{align}
These inequalities are also satisfied for the extrinsic noise $\sigma_{ex}^2$.

From Eq. (\ref{the decomposition of the mutual information into entropies}), when $\partial H(Y)/\partial \sigma_{in}^2>\partial H(Y|X)/\partial \sigma_{in}^2$, $I(X;Y)$ increases with $\sigma_{in}^2$. However, when $\partial H(Y)/\partial \sigma_{in}^2<\partial H(Y|X)/\partial \sigma_{in}^2$, $I(X;Y)$ decreases with $\sigma_{in}^2$. Therefore, we can interpret $\partial H(Y)/\partial \sigma_{in}^2$ as the beneficial effect of noise, acting to increase the transmitted information, and $\partial H(Y|X)/\partial \sigma_{in}^2$ as the harmful effect of noise, acting to decrease the transmitted information. Moreover, from Eq. (\ref{second derivative of H(Y) and H(Y|X)}), both effects of noise decrease as the intrinsic noise $\sigma_{in}^2$ increases.  

From Eqs. (\ref{the decomposition of the mutual information into entropies}) and (\ref{second derivative of H(Y) and H(Y|X)}), when $\partial^2 I(X;Y)/(\partial\sigma_{in}^2)^2>0$ (Fig. \ref{threshold model SR}(f)(blue)), the following is satisfied:

\begin{align}
	&\left|\frac{\partial^2 H(Y)}{\left(\partial \sigma_{in}^2\right)^2}\right|<\left|\frac{\partial^2 H(Y|X)}{\left(\partial \sigma_{in}^2\right)^2}\right|
\end{align}
Therefore, when $\partial^2 I(X;Y)/(\partial\sigma_{in}^2)^2>0$ (Fig. \ref{threshold model SR}(f)(blue)), the change in the harmful effect of noise $\partial H(Y|X)/\partial \sigma_{in}^2$ is greater than that in the beneficial effect of noise $\partial H(Y)/\partial \sigma_{in}^2$. Thus, we focus on the change in the harmful effect of noise $\partial H(Y|X)/\partial \sigma_{in}^2$. As mentioned above, from Eq. (\ref{second derivative of H(Y) and H(Y|X)}), the harmful effect of noise $\partial H(Y|X)/\partial \sigma_{in}^2$ decreases as the intrinsic noise $\sigma_{in}^2$ increases.  Therefore, when $\sigma_{in}^2$ is larger, the decrease in transmitted information as a result of noise is smaller. Thus, the robustness against extrinsic noise increases at higher levels of intrinsic noise.

In contrast, when $\partial^2 I(X;Y)/(\partial\sigma_{in}^2)^2<0$ (Fig. \ref{threshold model SR}(f)(green)), the following is satisfied:

\begin{align}
	&\left|\frac{\partial^2 H(Y)}{\left(\partial \sigma_{in}^2\right)^2}\right|>\left|\frac{\partial^2 H(Y|X)}{\left(\partial \sigma_{in}^2\right)^2}\right|
\end{align}
Therefore, when $\partial^2 I(X;Y)/(\partial\sigma_{in}^2)^2<0$ (Fig. \ref{threshold model SR}(f)(green)), the change in the beneficial effect of noise $\partial H(Y)/\partial \sigma_{in}^2$ is greater than that in the harmful effect of noise $\partial H(Y|X)/\partial \sigma_{in}^2$. Thus, we focus on the change in the beneficial effect of noise $\partial H(Y)/\partial \sigma_{in}^2$. From Eq. (\ref{second derivative of H(Y) and H(Y|X)}), the beneficial effect of noise $\partial H(Y)/\partial \sigma_{in}^2$ decreases as the intrinsic noise $\sigma_{in}^2$ increases. Therefore, when $\sigma_{in}^2$ is larger, the increase in transmitted information as a result of noise is smaller. Thus, the robustness against extrinsic noise decreases with the intrinsic noise.

\section{Discussion\label{section-Discussion}}
In this paper, we first analyzed the robustness against extrinsic noise in a linear model. We showed that the robustness against extrinsic noise increases with intrinsic noise. A threshold model in which the threshold is equal to the mean of the input was then analyzed, and we showed that the robustness against extrinsic noise again increases with the level of intrinsic noise. Therefore, INIR is realized in these models.  Moreover, we discovered that the intrinsic noise dependency of the robustness against extrinsic noise is stronger when the intrinsic noise is larger than the variance of the input in these models. 

This raises the question of whether the intrinsic noise is larger than the variance of the input in actual biological systems. In other words, is the robustness against the extrinsic noise strongly dependent on the intrinsic noise in actual biological systems? To address this question, we focus on the mutual information $I(X;Y)$ when the intrinsic noise is larger than the variance of the input, i.e., $\sigma_{in}^2>1$ in the linear model or $\sigma_{in}^2>2/(\pi\ln2)$ in the threshold model. In these cases, $I(X;Y)<0.5$ and $I(X;Y)<1-1/\sqrt{2}\approx0.29\cdots$ are satisfied in the linear model and in the threshold model, respectively. Therefore, we can estimate whether the robustness against the extrinsic noise is strongly dependent on the intrinsic noise from whether the mutual information is less than $0.3\sim0.5$ bits. Note that the input and noise distributions and input--output relations in actual biological systems are not exactly the same as those in our models. Therefore, this estimation is only approximate. With reference to several experiments that measure the mutual information between input molecules and cellular responses such as gene expressions, the range of the mutual information is $0.6\sim1.2$ bits \cite{Cheong2011,Uda2013,Selimkhanov2014,Voliotis2014,Suderman2017,Potter2017,Granados2017,Granados2018,Ruiz2018}. Therefore, the robustness against the extrinsic noise may not be  strongly dependent on the intrinsic noise at the cellular level. However, in computational studies analyzing a dendritic spine, which is a small part of a neuron, the range of the mutual information is $0.1\sim0.3$ bits \cite{Koumura2014,Fujii2017,Tottori2019}. Therefore, the robustness against the extrinsic noise may be  strongly dependent on the intrinsic noise in dendritic spines.

We showed that larger levels of intrinsic noise produce a higher degree of robustness against extrinsic noise in both the linear and threshold model when the threshold is equal to the mean of the input. However, larger values of intrinsic noise are not necessarily advantageous. This is because more intrinsic noise decreases the transmitted information, regardless of extrinsic noise. Therefore, in basic terms, cells may evolve to decrease the level of intrinsic noise as a means of increasing their transmitted information. However, intrinsic noise is often a significant component of biological systems \cite{Elowitz2002,Bar-Even2006,Newman2006,Taniguchi2010,Koumura2014,Fujii2017,Tottori2019}, and we only argue that the robustness against extrinsic noise is high in such cases. 

We further showed that SR appears in the threshold model when the threshold is much larger than the mean of the input. In this model, there is a region where the robustness against extrinsic noise decreases with the intrinsic noise, meaning that INIR is not always realized. However, in the threshold model with SR, the robustness against extrinsic noise and the transmitted information can be simultaneously increased by reducing the intrinsic noise. Therefore, SR solves the trade-off problem between the robustness against extrinsic noise and the transmitted information.

In artificial experimental settings,  SR has been observed in various biological systems, especially neuronal systems \cite{Douglass1993,Levin1996,Russell1999,Gluckman1996,Stacey2000,Stacey2001,Srebro1999,Mori2002,Kitajo2007,Stufflebeam2000,Tanaka2008,Ward2010}. Douglass et al. were the first to report the appearance of SR in biological systems, namely a sensory neuron in a crayfish tail fan \cite{Douglass1993}. Following this study, it has been reported that SR appears in the cercal sensory neurons of crickets \cite{Levin1996}, hippocampal slices of rats \cite{Gluckman1996,Stacey2000,Stacey2001}, electrical sensory organs of paddlefish \cite{Russell1999}, human vision \cite{Srebro1999,Mori2002,Kitajo2007}, and human audition \cite{Stufflebeam2000,Tanaka2008,Ward2010}. Moreover, the threshold response is observed in many biochemical reactions \cite{Ozbudak2004, Melen2005, Narula2012}, which may indicate the existence of SR. These reactions not only detect weak signals by utilizing noise, but also ensure robustness against additional noise. 

However, in natural environments, the threshold systems may not exhibit SR. This is because the mutual information is higher when the threshold is closer to the mean of the input. Indeed, in developmental processes, where the input--output relation between morphogen molecules and gene expressions is the threshold response, the number of cells with high gene expressions is similar to that with low gene expressions \cite{Gregor2007,Tkacik2008b}, which indicates that the threshold is close to the mean of the input. It has been reported that, even when the threshold is close to the mean of the input, SR appears if there are several threshold systems and the entire output is determined by the summation of individual outputs \cite{Stocks2000,Stocks2001}. Therefore, when the threshold  is equal to the mean of the input, SR does not appear in the information transmission by a single cell, but does appear in the information transmission by multiple cells \cite{Suderman2017}. For example, tissues and organs composed of multiple cells respond to the same input with multiple cells. Therefore, if individual cells exhibit a nonlinear response such as a threshold response, SR appears regardless of the relation between the input and the threshold. Indeed, several theoretical studies suggest that if multiple neurons transmit information, SR will even appear when the mean of the input is equal to the threshold \cite{Collins1995,Pei1996a,Chialvo1997,Stocks2001a}.

In this paper, we discussed SR in a static system, but SR can also appear in dynamical systems \cite{Gammaitoni1998}. Moreover, the transmitted information is conventionally quantified by the signal-to-noise ratio (SNR) in dynamical systems, rather than by the mutual information \cite{Gammaitoni1998}. However, even if it was quantified by other measures, the transmitted information and the robustness against the extrinsic noise could be simultaneously increased when SR appears, because there is always a region where the transmitted information is decreasing and convex upward with respect to the intrinsic noise. An example is given in Appendix \ref{appendix-sensitivity}.

In this paper, we assumed that the intrinsic noise and extrinsic noise obey Gaussian distributions. However, other types of distribution are often observed in biological systems \cite{Thattai2001,Shahrezaei2008,Taniguchi2010,Charlebois2011}. Moreover, we examined systems showing a linear response and a threshold response. However, smooth nonlinear response systems are often observed, as represented by Hill functions in biological systems.  To construct more a general theoretical framework and uncover the characteristics of INIR, analyses of these nonlinear systems should be considered.

Experimental verification of our theory is also necessary. Several experiments have measured the mutual information between extracellular molecules and cellular responses such as gene expressions \cite{Cheong2011,Uda2013,Selimkhanov2014,Voliotis2014,Suderman2017,Potter2017,Granados2017,Granados2018,Ruiz2018}. Furthermore, it is possible to generate extrinsic noise by changing the concentration of extracellular molecules from trial to trial. Therefore, we can measure the extrinsic noise dependency of the mutual information, i.e., the robustness against the extrinsic noise. Recent experimental studies have reported that the gene expression noise level can be changed without changing the gene expression mean by constructing synthetic gene circuits \cite{Farquhar2019}. This method enables the intrinsic noise to be modified without changing the input--output relation. Therefore, we could measure the intrinsic noise dependency of the robustness against extrinsic noise. These procedures would allow us to experimentally verify whether the robustness against extrinsic noise increases with the intrinsic noise in  a linear system, a threshold system without SR, and a threshold system with SR.

\begin{acknowledgments}
This work was supported by the Creation of Fundamental Technologies for Understanding and Control of Biosystem Dynamics, CREST (JPMJCR12W3), of the Japan Science and Technology Agency, and by the Japan Society for the Promotion of Science (JSPS), KAKENHI Grant Numbers 17H06300, 17H6299, 18H03979 and 19K22860. M.F. was funded by the JSPS, KAKENHI Grant Numbers 16K12508 and 19K20382.
\end{acknowledgments}

\appendix
\section{Robustness function $\delta_{q}(\sigma_{in}^2)$ in the linear model}
\subsection{Derivation of Eq. (\ref{LRM-the robustness function})\label{appendix-LRM-robustness-1}}
From Eqs. (\ref{the definition of the robustness function}) and (\ref{LRM-MI}), 

\begin{align}
	\frac{\frac{1}{2}\log_{2}\left(1+\frac{1}{\delta_{q}(\sigma_{in}^2)+\sigma_{in}^2}\right)}{\frac{1}{2}\log_{2}\left(1+\frac{1}{\sigma_{in}^2}\right)}&=q
	\label{the definition of the robustness function-and-LRM-MI}
\end{align}
is obtained. From Eq. (\ref{the definition of the robustness function-and-LRM-MI}), we have

\begin{align}
	\delta_{q}(\sigma_{in}^2)&=\frac{1}{\left(1+\sigma_{in}^{-2}\right)^{q}-1}-\sigma_{in}^2
\end{align}
which is the same as Eq. (\ref{LRM-the robustness function}).

\subsection{Derivation of Eq. (\ref{LRM-the robustness function-approx})\label{appendix-LRM-robustness-2}}
When $\sigma_{in}^2\gg1$, Eq. (\ref{LRM-the robustness function}) can be approximated as follows:

\begin{align}
	\delta_{q}(\sigma_{in}^2)&=\frac{1}{\left(1+\sigma_{in}^{-2}\right)^{q}-1}-\sigma_{in}^2\nonumber\\
	&\approx\frac{1}{1+q\sigma_{in}^{-2}-1}-\sigma_{in}^2
	=\frac{1-q}{q}\sigma_{in}^{2}
	\label{Apx-LRM-the robustness function-first-approx}
\end{align}

When $\sigma_{in}^2\ll1$, Eq. (\ref{LRM-the robustness function}) can be approximated as:

\begin{align}
	\delta_{q}(\sigma_{in}^2)&=\frac{1}{\left(1+\sigma_{in}^{-2}\right)^{q}-1}-\sigma_{in}^2\nonumber\\
	&\approx\frac{1}{\sigma_{in}^{-2q}-1}-\sigma_{in}^2
\end{align}
When $\sigma_{in}^{2q}\ll1$, 

\begin{align}
	\delta_{q}(\sigma_{in}^2)&\approx\sigma_{in}^{2q}-\sigma_{in}^2
\end{align}
When $\sigma_{in}^{2q}\gg\sigma_{in}^2$,

\begin{align}
	\delta_{q}(\sigma_{in}^2)&\approx\sigma_{in}^{2q}
	\label{Apx-LRM-the robustness function-second-approx}
\end{align}

Therefore, from Eqs. (\ref{Apx-LRM-the robustness function-first-approx}) and (\ref{Apx-LRM-the robustness function-second-approx}), we find that

\begin{align}
	\delta_{q}(\sigma_{in}^2)
	\propto\begin{cases}
		\sigma_{in}^{2q} & (\sigma_{in}^2\ll1)\\
		\sigma_{in}^2 & (\sigma_{in}^2\gg1)		
	\end{cases}
\end{align}
which is the same as Eq. (\ref{LRM-the robustness function-approx}). More precisely, $\delta_{q}(\sigma_{in}^2)\propto\sigma_{in}^{2q}$ is not satisfied when $\sigma_{in}^2\ll1$, but holds when $\sigma_{in}^{2}\ll\sigma_{in}^{2q}$.

\section{Mutual information $I(X;Y)$ in the threshold model ($\theta=0$)}
\subsection{Derivation of Eq. (\ref{TRM-theta0-MI})\label{appendix-TRM-MI-1}}
In the threshold model, the probability density function of $X$, $p(x)$, the probability mass function of $Y$ given by $X$, $p(y|x)$, and the probability mass function of $Y$, $p(y)$, are given by the following: 

\begin{align}
	p(x)&=N(x|0,1)\label{appendix-p(x)}\\
	p(y=1|x)&=\Phi\left(\frac{x-\theta}{\sqrt{\sigma_{ex}^2+\sigma_{in}^2}}\right)\label{appendix-p(y=1|x)}\\
	p(y=0|x)&=1-p(y=1|x)\label{appendix-p(y=0|x)}\\
	p(y=1)&=\Phi\left(\frac{-\theta}{\sqrt{1+\sigma_{ex}^2+\sigma_{in}^2}}\right)\\
	p(y=0)&=1-p(y=0|x)
\end{align}
where $\Phi(x)=\int_{-\infty}^{x}N(z|0,1)dz$ is the cumulative distribution function of the standard normal distribution.

The mutual information $I(X;Y)$ is calculated as follows:

\begin{align}
	I(X;Y)&=H(Y)-H(Y|X)
	\label{appendix-decomposition of I(X;Y)}
\end{align}
where

\begin{align}
	H(Y)&=-\sum_{y\in\{0,1\}}p(y)\log_{2}p(y)\\
	H(Y|X)&=-\int_{-\infty}^{\infty}p(x)\sum_{y\in\{0,1\}}p(y|x)\log_{2}p(y|x)dx
\end{align}
$H(Y)$ is the entropy of $Y$, and $H(Y|X)$ is the entropy of $Y$ given $X$.

When $\theta=0$, $p(y=1)=p(y=0)=1/2$. Therefore, 

\begin{align}
	H(Y)&=-2\cdot\frac{1}{2}\log_{2}\frac{1}{2}=1
	\label{appendix-H(Y)}
\end{align}
$H(Y|X)$ cannot be calculated analytically. Therefore, we calculate the approximate solution of $H(Y|X)$ as follows. First, $H(Y|X)$ is calculated as:

\begin{align}
	H(Y|X)&=\int_{-\infty}^{\infty}p(x)H(Y|X=x)dx
	\label{decomposition of H(Y|X)}
\end{align}
where 

\begin{align}
	H(Y|X=x)&=-\sum_{y\in\{0,1\}}p(y|x)\log_{2}p(y|x)
\end{align}
and $H(Y|X=x)$ is the entropy of $Y$ given $X=x$.

\begin{figure}[btp]
\begin{center}
	\includegraphics[width=75mm]{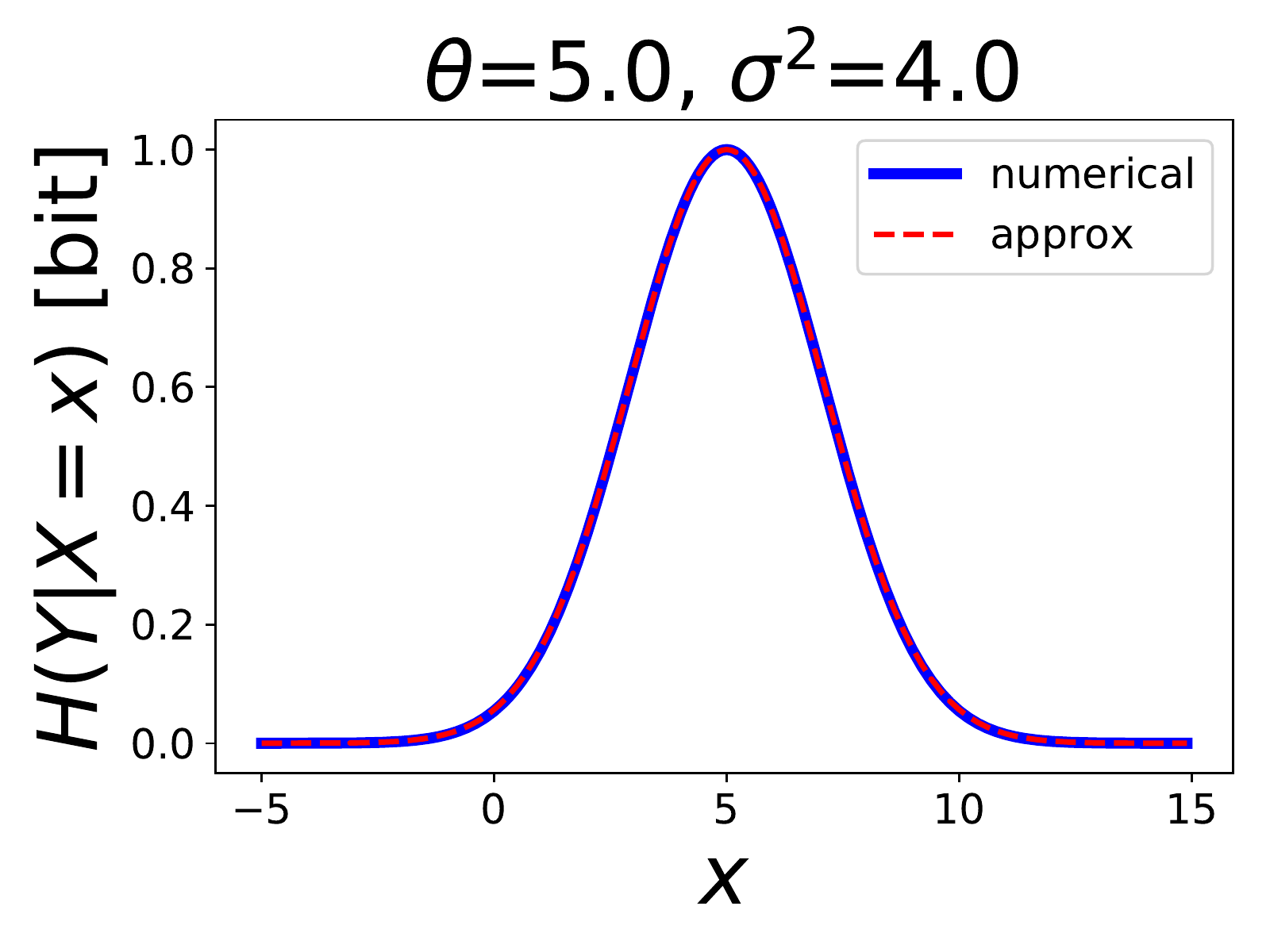}\\[-1pc]
	 \caption{
	$H(Y|X=x)$ in the threshold model (blue) and the Gaussian function (red). We define $\sigma^2:=\sigma_{ex}^2+\sigma_{in}^2$.
	}
	\label{H(Y|X=x) in the threshold model and the gaussian function}
\end{center}
\end{figure}

As $H(Y|X=x)$ is similar to the Gaussian function $a\exp(-(x-b)^2/(2c^2))$ (Fig. \ref{H(Y|X=x) in the threshold model and the gaussian function}), we approximate $H(Y|X=x)$ by the Gaussian function $a\exp(-(x-b)^2/(2c^2))$, which gives the following:

\begin{align}
	H(Y|X=x)&\approx\exp\left(-\frac{(x-\theta)^2}{(\pi\ln 2)(\sigma_{ex}^2+\sigma_{in}^2)}\right)
	\label{laplace-approximation}
\end{align}
(see Appendix \ref{appendix-TRM-MI-2}).

From Eqs. (\ref{appendix-p(x)}), (\ref{decomposition of H(Y|X)}), and (\ref{laplace-approximation}), $H(Y|X)$ can be calculated as follows:

\begin{align}
	H(Y|X)\approx&\frac{1}{\sqrt{\frac{2}{(\pi\ln2)}\left(\sigma_{ex}^2+\sigma_{in}^2\right)^{-1}+1}}\nonumber\\
	&\times\exp\left(-\frac{\theta^2}{(\pi\ln2)\left(\frac{2}{(\pi\ln2)}+\sigma_{ex}^2+\sigma_{in}^2\right)}\right)
	\label{appendix-H(Y|X)}
\end{align}
Therefore, from Eqs. (\ref{appendix-decomposition of I(X;Y)}), (\ref{appendix-H(Y)}), and (\ref{appendix-H(Y|X)}), the approximate solution of $I(X;Y)$ in the threshold model ($\theta=0$) is given by the following:

\begin{align}
	I(X;Y)&\approx1-\frac{1}{\sqrt{\frac{2}{(\pi\ln2)}\left(\sigma_{ex}^2+\sigma_{in}^2\right)^{-1}+1}}
\end{align}
which is the same as Eq. (\ref{TRM-theta0-MI}).

\subsection{Derivation of Eq. (\ref{laplace-approximation})\label{appendix-TRM-MI-2}}
In this subsection, we explain how to approximate $H(Y|X=x)$ in the threshold model as the Gaussian function $a\exp(-(x-b)^2/(2c^2))$. For simplicity, we define $\sigma^2:=\sigma_{ex}^2+\sigma_{in}^2$.

First, we derive $b$, which is the value of $x$ that satisfies $\partial H(Y|X=x)/\partial x=0$. $\partial H(Y|X=x)/\partial x$ is given by:

\begin{align}
	\frac{\partial H(Y|X=x)}{\partial x}&=-\frac{1}{\ln 2}\sum_{y\in\{0,1\}}\left\{p'(y|x)\ln p(y|x)+p'(y|x)\right\}
\end{align}
where $p'(y|x)$ represents $\partial p(y|x)/\partial x$, which is given by:

\begin{align}
	p'(y=1|x)&=\frac{\partial}{\partial x}\Phi\left(\frac{x-\theta}{\sigma}\right)
	=\frac{1}{\sigma}N\left(\left.\frac{x-\theta}{\sigma}\right|0,1\right)\nonumber\\
	p'(y=0|x)&=-p'(y=1|x)\label{Derivation of Eq. (laplace-approximation)-p'(y=0|x)}
\end{align}
Note that we used $(d/dx)\int_{a}^{x}f(z)dz=f(x)$, where $a$ is constant. From Eq. (\ref{Derivation of Eq. (laplace-approximation)-p'(y=0|x)}), $\partial H(Y|X=x)/\partial x$ is calculated as follows:

\begin{align}
	\frac{\partial H(Y|X=x)}{\partial x}=&-\frac{1}{\ln 2}p'(y=1|x)\nonumber\\
	&\times\left\{\ln p(y=1|x)-\ln p(y=0|x)\right\}
\end{align}
From $p'(y=1|x)>0$, when $\partial H(Y|X=x)/\partial x=0$, 

\begin{align}
	p(y=1|x)=p(y=0|x)
\end{align}
is satisfied. Accordingly, $b=\theta$.

Next, we derive $c$, which satisfies 

\begin{align}
	\frac{1}{c^2}&=-\left.\frac{\partial^2}{\partial x^2}\ln H(Y|X=x)\right|_{x=\theta}
	\label{definition of c}
\end{align}
$\partial \ln H(Y|X=x)/\partial x$ is given by the following: 

\begin{align}
	&\frac{\partial}{\partial x}\ln H(Y|X=x)=\frac{H'(Y|X=x)}{H(Y|X=x)}
\end{align}
Therefore, $\partial^2 \ln H(Y|X=x)/\partial x^2$ is given by: 

\begin{align}
	&\frac{\partial^2}{\partial x^2}\ln H(Y|X=x)\nonumber\\
	=&\frac{H''(Y|X=x)H(Y|X=x)-H'(Y|X=x)^2}{H(Y|X=x)^2}
\end{align}
When $x=\theta$, we have 

\begin{align}
	H(Y|X=x)&=1\\
	H'(Y|X=x)&=0\\
	H''(Y|X=x)&=-\frac{2}{(\pi\ln 2)\sigma^2}
\end{align}
Therefore, $\partial^2 \ln H(Y|X=x)/\partial x^2|_{x=\theta}$ can be written as: 

\begin{align}
	\left.\frac{\partial^2}{\partial x^2}\ln H(Y|X=x)\right|_{x=\theta}=-\frac{2}{(\pi\ln 2)\sigma^2}
\end{align}
Therefore, from Eq. (\ref{definition of c}), we have that

\begin{align}
	c^2=\frac{(\pi\ln 2)\sigma^2}{2}
\end{align}

Finally, we derive $a$. $a$ is given by $\left.H(Y|X=x)\right|_{x=\theta}$. Therefore, $a=1$.

Thus, the approximate solution of $H(Y|X=x)$ is given by: 

\begin{align}
	H(Y|X=x)&\approx\exp\left(-\frac{(x-\theta)^2}{(\pi\ln 2)\sigma^2}\right)
\end{align}
which is the same as Eq. (\ref{laplace-approximation}).

\section{Robustness function $\delta_{q}(\sigma_{in}^2)$ in the threshold model ($\theta=0$)}
\subsection{Derivation of Eq. (\ref{TRM-the robustness function})\label{appendix-TRM-robustness-1}}
From Eqs. (\ref{the definition of the robustness function}) and (\ref{TRM-theta0-MI}), 

\begin{align}
	1-\frac{1}{\sqrt{\frac{2}{(\pi\ln2)}\left(\delta_{q}(\sigma_{in}^2)+\sigma_{in}^2\right)^{-1}+1}}&\approx qI_{0}
	\label{Derivation of Eq. (TRM-the robustness function)-C1}
\end{align}
is satisfied, where 

\begin{align}
	I_{0}=1-\frac{1}{\sqrt{\frac{2}{(\pi\ln2)}\sigma_{in}^{-2}+1}}
\end{align}
From Eq. (\ref{Derivation of Eq. (TRM-the robustness function)-C1}), we have

\begin{align}
	\delta_{q}(\sigma_{in}^2)
	&\approx\frac{\left\{1-qI_{0}\right\}^2}{\left\{2-qI_{0}\right\}qI_{0}}\frac{2}{(\pi\ln2)}-\sigma_{in}^2
\end{align}
which is the same as Eq. (\ref{TRM-the robustness function}).

\subsection{Derivation of Eq. (\ref{TRM-the robustness function-approx})\label{appendix-TRM-robustness-2}}
When $\sigma_{in}^2\gg\frac{2}{\pi\ln2}$, Eq. (\ref{TRM-the robustness function-I0}) can be approximated as follows:

\begin{align}
	I_{0}&=1-\frac{1}{\sqrt{\frac{2}{(\pi\ln2)}\sigma_{in}^{-2}+1}}
	\approx\frac{1}{(\pi\ln2)}\sigma_{in}^{-2}
\end{align}
Therefore, Eq. (\ref{TRM-the robustness function}) can be approximated as:

\begin{align}
	\delta_{q}(\sigma_{in}^2)
	&\approx\frac{\left\{1-qI_{0}\right\}^2}{\left\{2-qI_{0}\right\}qI_{0}}\frac{2}{(\pi\ln2)}
	-\sigma_{in}^2\nonumber\\
	&\approx\frac{1}{\frac{2q}{(\pi\ln2)}\sigma_{in}^{-2}}\frac{2}{(\pi\ln2)}-\sigma_{in}^2
	=\left(\frac{1-q}{q}\right)\sigma_{in}^{2}
	\label{Apx-TRM-the robustness function-first-approx}
\end{align}

When $\sigma_{in}^2\ll\frac{2}{\pi\ln2}$, Eq. (\ref{TRM-the robustness function-I0}) can be approximated as follows:

\begin{align}
	I_{0}&=1-\frac{1}{\sqrt{\frac{2}{(\pi\ln2)}\sigma_{in}^{-2}+1}}
	\approx1-\sqrt{\frac{(\pi\ln2)}{2}}\sigma_{in}
\end{align}
Therefore, Eq. (\ref{TRM-the robustness function}) can be approximated as:

\begin{align}
	\delta_{q}(\sigma_{in}^2)
	&\approx\frac{\left\{1-qI_{0}\right\}^2}{\left\{2-qI_{0}\right\}qI_{0}}\frac{2}{(\pi\ln2)}
	-\sigma_{in}^2\nonumber\\
	&\approx\frac{\left\{\left(\frac{1-q}{q}\right)+\sqrt{\frac{(\pi\ln2)}{2}}\sigma_{in}\right\}^2}
	{\left(\frac{2-q}{q}\right)}\frac{2}{(\pi\ln2)}-\sigma_{in}^2
	\label{Apx-TRM-the robustness function-approx-eq0}
\end{align}
Rearranging this expression in terms of $\sigma_{in}$ gives: 

\begin{align}
	\delta_{q}(\sigma_{in}^2)&=\frac{4}{(\pi\ln2)}\frac{(1-q)}{(2-q)}
	\left\{\frac{(1-q)}{2q}\right.\nonumber\\
	&\ \ \ \ \ \ \left.+\sqrt{\frac{(\pi\ln2)}{2}}\sigma_{in}+\frac{(\pi\ln2)}{2}\sigma_{in}^2\right\}
\end{align}
From $\sigma_{in}^2\ll\frac{2}{\pi\ln2}$, the equation above can be approximated as follows:

\begin{align}
	\delta_{q}(\sigma_{in}^2)
	&=\frac{4}{(\pi\ln2)}\frac{(1-q)}{(2-q)}\left\{\frac{(1-q)}{2q}+\sqrt{\frac{(\pi\ln2)}{2}}\sigma_{in}\right\}
	\label{Apx-TRM-the robustness function-approx-eq1}
\end{align}

When $\frac{(1-q)^2}{4q^2}\frac{2}{\pi\ln2}\ll\sigma_{in}^2\ll\frac{2}{\pi\ln2}$, Eq. (\ref{Apx-TRM-the robustness function-approx-eq1}) can be approximated as:

\begin{align}
	\delta_{q}(\sigma_{in}^2)&\approx2\sqrt{\frac{2}{(\pi\ln2)}}\frac{(1-q)}{(2-q)}\sigma_{in}
	\label{Apx-TRM-the robustness function-second-approx}
\end{align}
When $\sigma_{in}^2\ll\min\left(\frac{(1-q)^2}{4q^2}\frac{2}{\pi\ln2},\frac{2}{\pi\ln2}\right)$, Eq. (\ref{Apx-TRM-the robustness function-approx-eq1}) can be approximated as:

\begin{align}
	\delta_{q}(\sigma_{in}^2)&\approx\frac{2}{(\pi\ln2)}\frac{(1-q)^2}{q(2-q)}
	\label{Apx-TRM-the robustness function-third-approx}
\end{align}

Therefore, from Eqs. (\ref{Apx-TRM-the robustness function-first-approx}), (\ref{Apx-TRM-the robustness function-second-approx}), and (\ref{Apx-TRM-the robustness function-third-approx}), we obtain

\begin{align}
	\delta_{q}(\sigma_{in}^2)
	&\approx\frac{\left\{1-qI_{0}\right\}^2}{\left\{2-qI_{0}\right\}qI_{0}}\frac{2}{(\pi\ln2)}-\sigma_{in}^2\nonumber\\
	&\propto\begin{cases}
		\sigma_{in}^{2} & \left(\sigma_{in}^2\gg\frac{2}{\pi\ln2}\right)\\
		\sigma_{in} & \left(\frac{(1-q)^2}{4q^2}\frac{2}{\pi\ln2}\ll\sigma_{in}^2\ll\frac{2}{\pi\ln2}\right)\\
		{\rm const} & \left(\sigma_{in}^2\ll\min\left(\frac{(1-q)^2}{4q^2}\frac{2}{\pi\ln2},\frac{2}{\pi\ln2}\right)\right)
	\end{cases}
\end{align}
which is the same as Eq. (\ref{TRM-the robustness function-approx}).

\section{Mutual information $I(X;Y)$ in the threshold model ($\theta\neq0$)}
\subsection{Derivation of Eq. (\ref{TRM-SR-MI})\label{appendix-TRM-SR-MI-1}}
The mutual information $I(X;Y)$ is calculated as follows:

\begin{align}
	I(X;Y)&=H(Y)-H(Y|X)
\end{align}
where

\begin{align}
	H(Y)&=-\sum_{y\in\{0,1\}}p(y)\log_{2}p(y)\\
	H(Y|X)&=-\int_{-\infty}^{\infty}p(x)\sum_{y\in\{0,1\}}p(y|x)\log_{2}p(y|x)dx
\end{align}
$H(Y)$ is the entropy of $Y$, and $H(Y|X)$ is the entropy of $Y$ given $X$.

In Appendix \ref{appendix-TRM-MI-1}, by approximating $H(Y|X=x)$ using a Gaussian function, we obtained the approximate solution of $H(Y|X)$ as:

\begin{align}
	H(Y|X)\approx&\frac{1}{\sqrt{\frac{2}{(\pi\ln2)}\left(\sigma_{ex}^2+\sigma_{in}^2\right)^{-1}+1}}\nonumber\\
	&\times\exp\left(-\frac{\theta^2}{(\pi\ln2)\left(\frac{2}{(\pi\ln2)}+\sigma_{ex}^2+\sigma_{in}^2\right)}\right)
\end{align}

In the same way, we can approximate $H(Y)$ using a Gaussian function to give:

\begin{align}
	H(Y)&\approx\exp\left(-\frac{\theta^2}{(\pi\ln 2)\left(1+\sigma_{ex}^2+\sigma_{in}^2\right)}\right)
\end{align}

Therefore, $I(X;Y)$ can be approximated as follows:

\begin{align}
	I(X;Y)
	\approx&\exp\left(-\frac{\theta^2}{(\pi\ln 2)\left(1+\sigma_{ex}^2+\sigma_{in}^2\right)}\right)\nonumber\\
	&-\frac{1}{\sqrt{\frac{2}{(\pi\ln2)}\left(\sigma_{ex}^2+\sigma_{in}^{2}\right)^{-1}+1}}\nonumber\\
	&\times\exp\left(-\frac{\theta^2}{(\pi\ln2)\left(\frac{2}{(\pi\ln2)}+\sigma_{ex}^2+\sigma_{in}^{2}\right)}\right)
\end{align}

Here, $\frac{2}{\pi\ln2}=0.91844\cdots\approx1$. Therefore, $H(Y)$ can be approximated as:

\begin{align}
	H(Y)&\approx\exp\left(-\frac{\theta^2}{(\pi\ln 2)\left(\frac{2}{(\pi\ln2)}+\sigma_{ex}^2+\sigma_{in}^2\right)}\right)
\end{align}
Thus, the mutual information $I(X;Y)$ can be approximated as follows:

\begin{align}
	I(X;Y)
	&\approx\left(1-\frac{1}{\sqrt{\frac{2}{(\pi\ln2)}\left(\sigma_{ex}^2+\sigma_{in}^{2}\right)^{-1}+1}}\right)\nonumber\\
	&\ \ \ \times\exp\left(-\frac{\theta^2}{(\pi\ln2)\left(\frac{2}{(\pi\ln2)}+\sigma_{ex}^2+\sigma_{in}^{2}\right)}\right)
\end{align}
which is the same as Eq. (\ref{TRM-SR-MI}).

\subsection{Derivation of the condition under which SR appears: $|\theta|>2$\label{appendix-TRM-SR-MI-2}}
We can derive the condition for the appearance of SR from the approximate solution of the mutual information $I(X;Y)$. For simplicity, we consider the case $\sigma_{ex}^2=0$. Note that we can easily expand this to the case $\sigma_{ex}^2\geq0$. 

When SR appears, the mutual information $I(X;Y)$ has a local maximum with respect to the intrinsic noise $\sigma_{in}^2$. Therefore, when SR appears, there is some value of $\sigma_{in}^2$ for which $\partial I(X;Y)/\partial \sigma_{in}^2=0$. Therefore, we derive the condition that ensures $\partial I(X;Y)/\partial \sigma_{in}^2=0$.

$\partial I(X;Y)/\partial \sigma_{in}^2$ is given by the following:

\begin{align}
	&\frac{\partial I(X;Y)}{\partial\sigma_{in}^2}\nonumber\\
	\approx&-\left\{\left(\frac{2}{(\pi\ln2)}\sigma_{in}^{-2}+1\right)-\theta^2\sqrt{\frac{2}{(\pi\ln2)}\sigma_{in}^{-2}+1}+\theta^2\right\}\nonumber\\
	&\times\frac{\sigma_{in}^{-4}\exp\left(-\frac{\theta^2}{(\pi\ln2)\left(\frac{2}{(\pi\ln2)}+\sigma_{in}^2\right)}\right)}
	{(\pi\ln2)\left(\frac{2}{(\pi\ln2)}\sigma_{in}^{-2}+1\right)^{\frac{5}{2}}}
\end{align}
Therefore, when $\partial I(X;Y)/\partial \sigma_{in}^2=0$ holds, 

\begin{align}
	\left(\frac{2}{(\pi\ln2)}\sigma_{in}^{-2}+1\right)-\theta^2\sqrt{\frac{2}{(\pi\ln2)}\sigma_{in}^{-2}+1}+\theta^2&=0
\end{align}
is satisfied. Defining $A=\sqrt{\frac{2}{(\pi\ln2)}\sigma_{in}^{-2}+1}$, 

\begin{align}
	A^2-\theta^2A+\theta^2=0
	\label{quadratic equation of SR}
\end{align}
Here, the discriminant $D$ is given by:

\begin{align}
	D:=\theta^4-4\theta^2=\theta^2\left(\theta^2-4\right)
\end{align}
As $A$ is a real number, $D$ must satisfy $D\geq0$. Therefore, 

\begin{align}
	\theta=0,\ \left|\theta\right|\geq2
\end{align}
is the necessary condition for $A$ to satisfy Eq. (\ref{quadratic equation of SR}).

When $\theta=0$, the solution of Eq. (\ref{quadratic equation of SR}) is $A=0$. This is contrary to $A>1$. Therefore, when $\theta=0$, there is no value of $\sigma_{in}^2$ that satisfies Eq. (\ref{quadratic equation of SR}), and SR does not appear.

When $\left|\theta\right|\geq2$, the solutions of Eq. (\ref{quadratic equation of SR}) are given by the following:

\begin{align}
	A_{\pm}=\frac{\theta^2\pm\sqrt{\theta^4-4\theta^2}}{2}=\frac{\theta^2}{2}\left\{1\pm\sqrt{1-\frac{4}{\theta^2}}\right\}
\end{align}
From simple calculations, we can confirm that $A_{\pm}$ satisfies $A_{\pm}>1$. Therefore, $\left|\theta\right|>2$ is the necessary and sufficient condition for the appearance of SR.

\begin{figure}[btp]
\begin{center}
	\includegraphics[width=75mm]{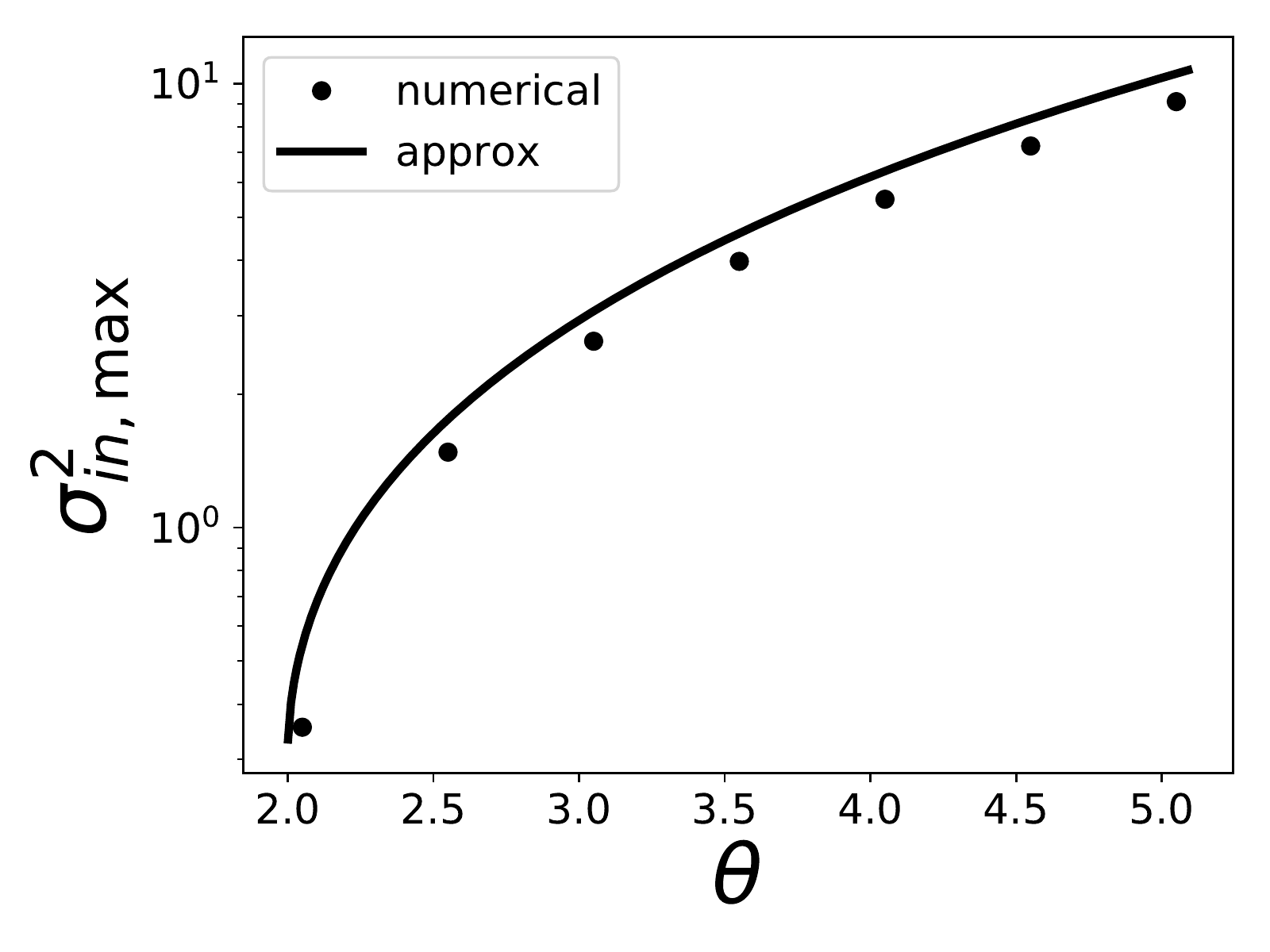}\\[-1pc]
	 \caption{
	$\sigma_{in,\max}^2$ against the threshold $\theta$.
	}
	\label{fig-sigma_in2_max}
\end{center}
\end{figure}

Furthermore, we can derive the intrinsic noise $\sigma_{in}^2$ that gives the local maximum of $I(X;Y)$, defined as $\sigma_{in,\max}^2$.

Defining $\sigma_{in,\pm}^2$ as the intrinsic noise $\sigma_{in}^2$ corresponding to $A=A_{\pm}$, 

\begin{align}
	\sigma_{in,\pm}^{2}&=\frac{2}{(\pi\ln2)}\left[A_{\pm}^2-1\right]^{-1}
\end{align}
As $\sigma_{in}^{2}=\sigma_{in,\pm}^{2}$ satisfies $\partial I(X;Y)/\partial \sigma_{in}^2=0$, $\sigma_{in,\pm}^{2}$ gives the local maximum or local minimum of $I(X;Y)$.

From numerical calculations, it can be clarified that $\sigma_{in,+}^{2}$ gives the local minimum of $I(X;Y)$ and $\sigma_{in,-}^{2}$ gives the local maximum of $I(X;Y)$. Therefore, 

\begin{align}
	\sigma_{in,\max}^2&=\sigma_{in,-}^2:=\frac{2}{(\pi\ln2)}\left[A_{-}^2-1\right]^{-1}
\end{align}
is satisfied. This approximate solution matches the numerical solution well (Fig. \ref{fig-sigma_in2_max}).

\section{Robustness function $\delta_{q}(\sigma_{in}^2)$ defined not by mutual information but by the other measure\label{appendix-sensitivity}}
\begin{figure}[btp]
\begin{center}
	\includegraphics[width=75mm]{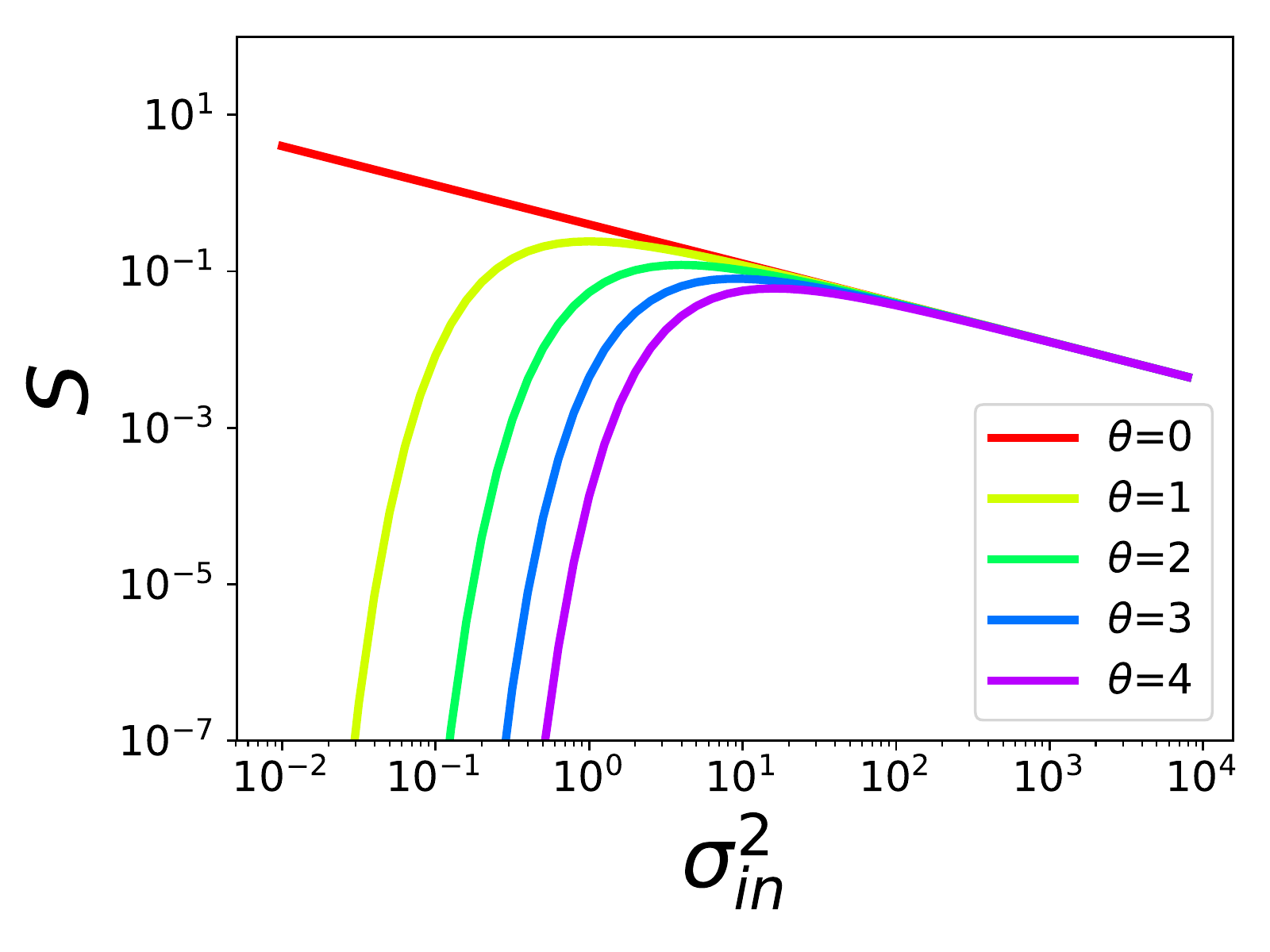}
	 \caption{
	 Sensitivity $S$ against the intrinsic noise $\sigma_{in}^2$ at $\sigma_{ex}^2=0$ in the threshold model. 
	 }
	\label{fig-sensitivity}
\end{center}
\end{figure}

\begin{figure}[btp]
\begin{center}
	(a)\hspace{-3.0mm}
	\begin{minipage}[t][][b]{40mm}
		\includegraphics[width=40mm,height=35mm]{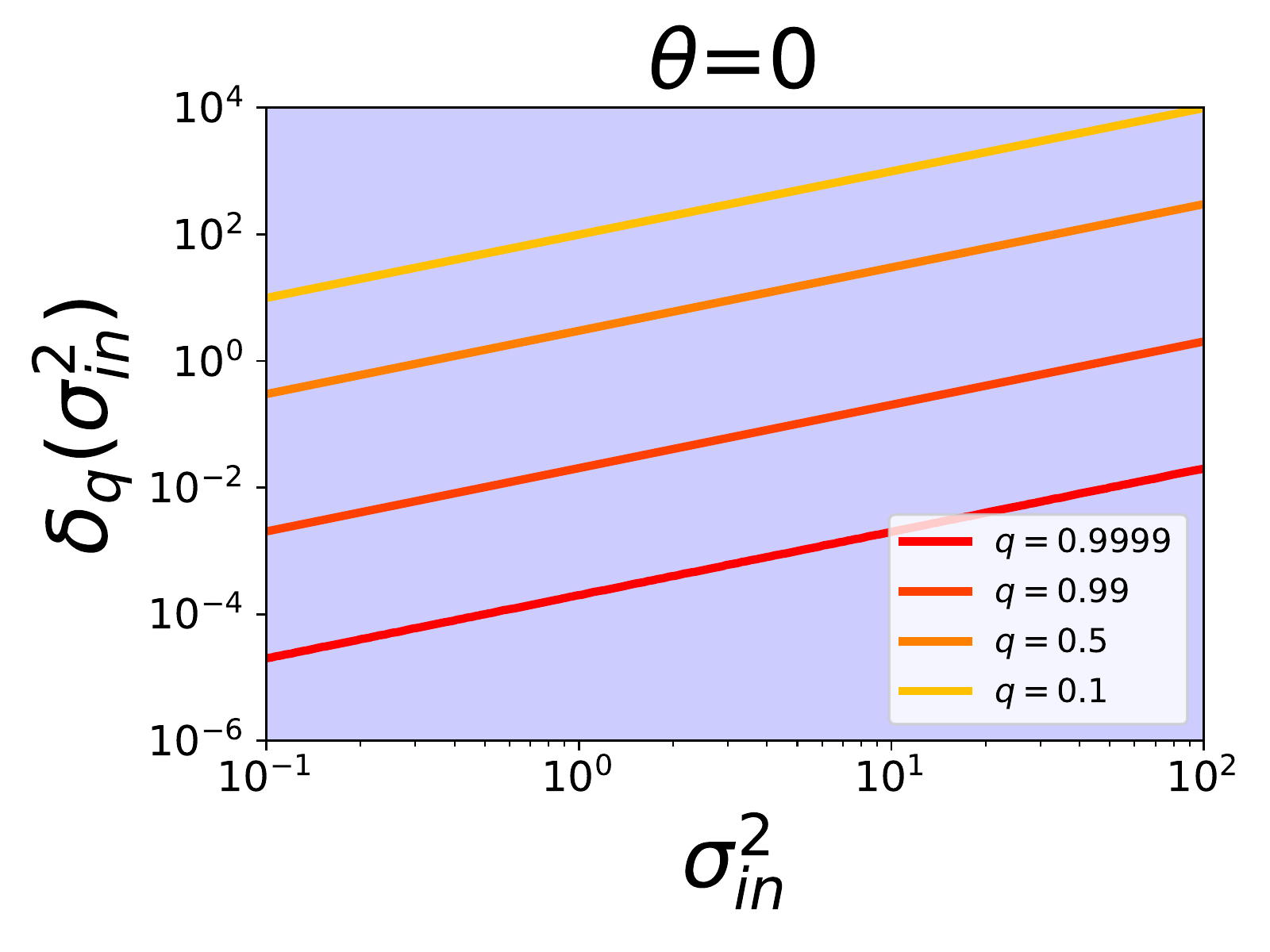}
	\end{minipage}
	(b)\hspace{-3.0mm}
	\begin{minipage}[t][][b]{40mm}
		\includegraphics[width=40mm,height=35mm]{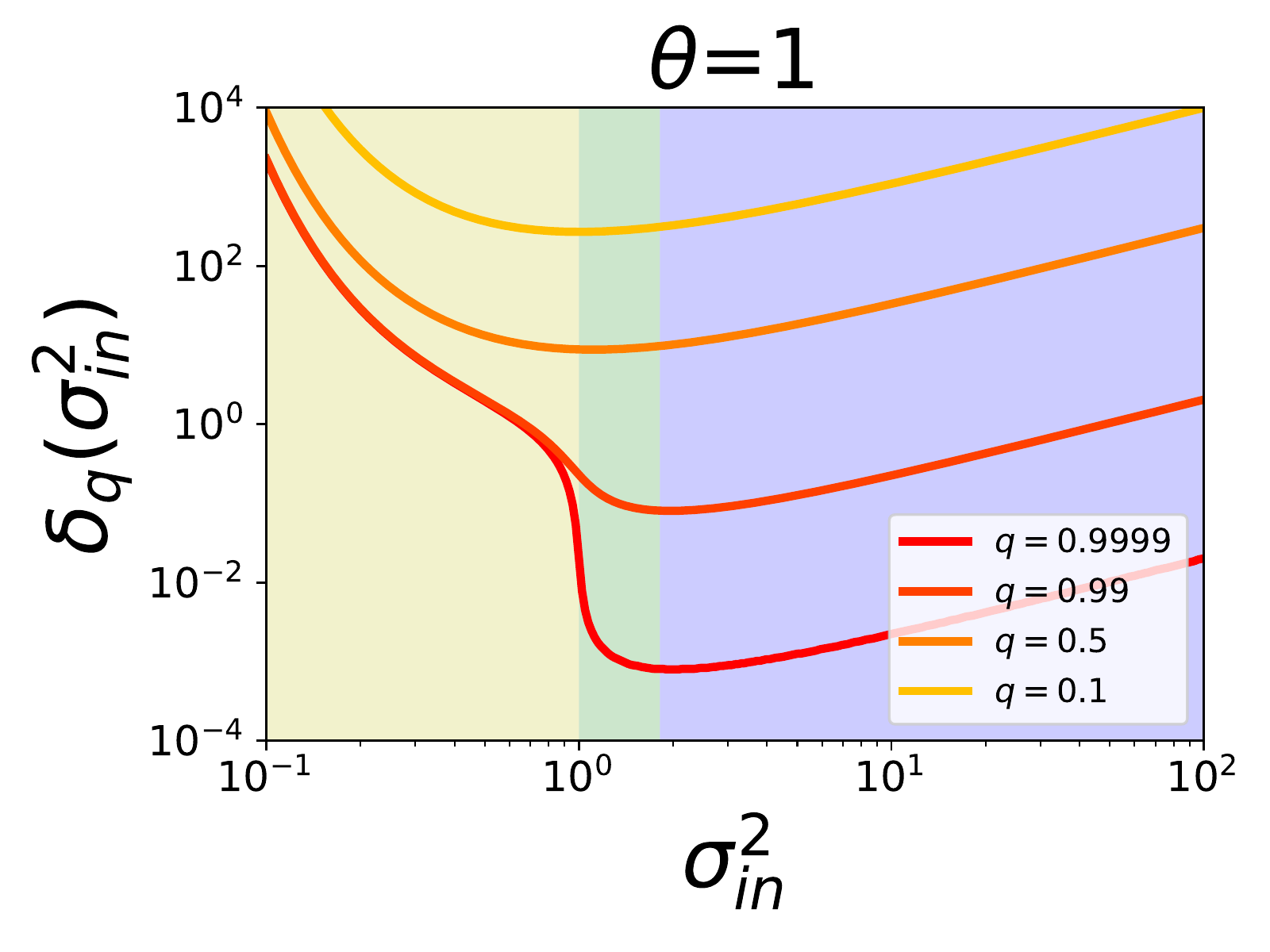}
	\end{minipage}\\
	(c)\hspace{-3.0mm}
	\begin{minipage}[t][][b]{40mm}
		\includegraphics[width=40mm,height=35mm]{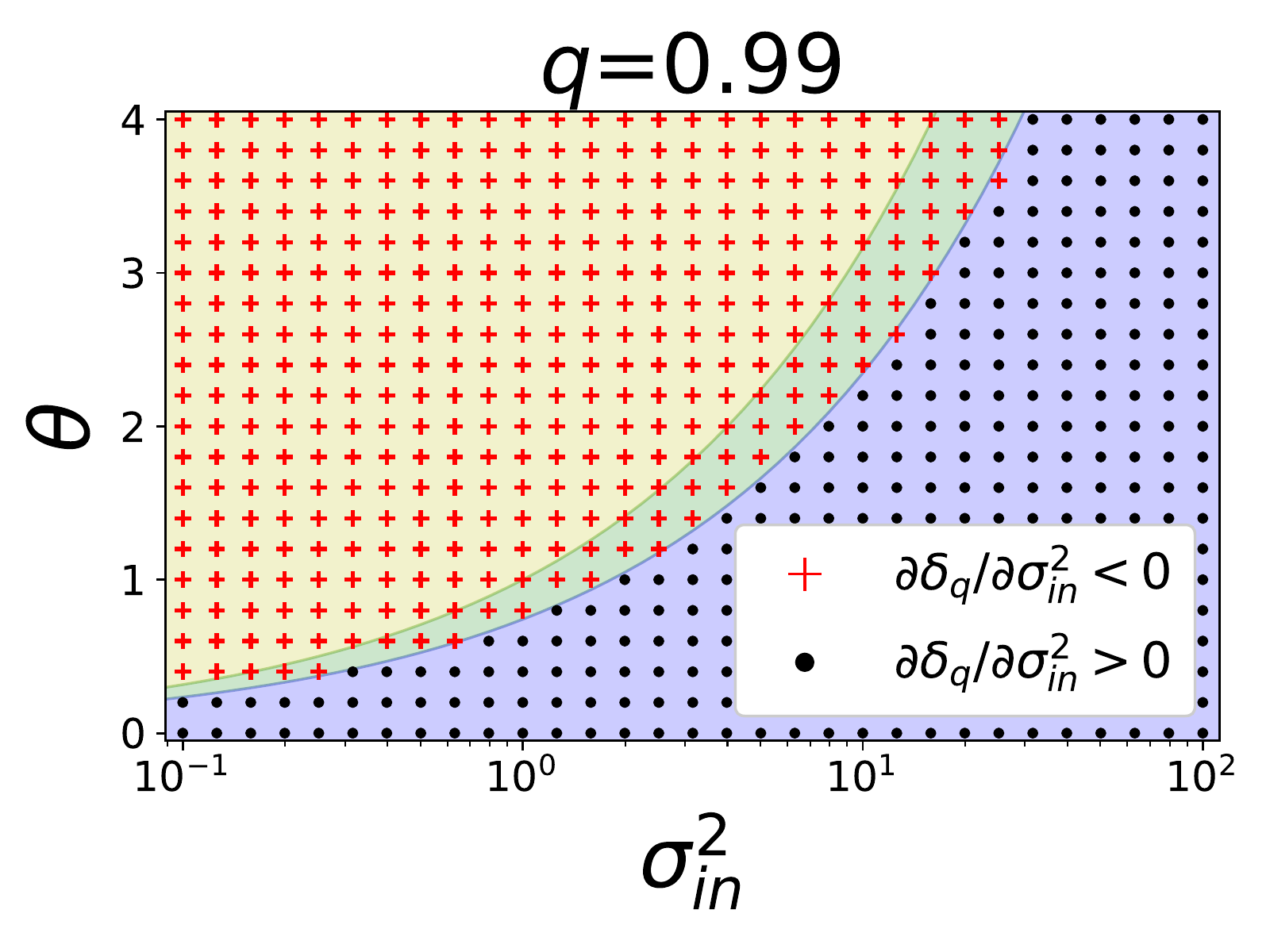}
	\end{minipage}
	(d)\hspace{-3.0mm}
	\begin{minipage}[t][][b]{40mm}
		\includegraphics[width=40mm,height=35mm]{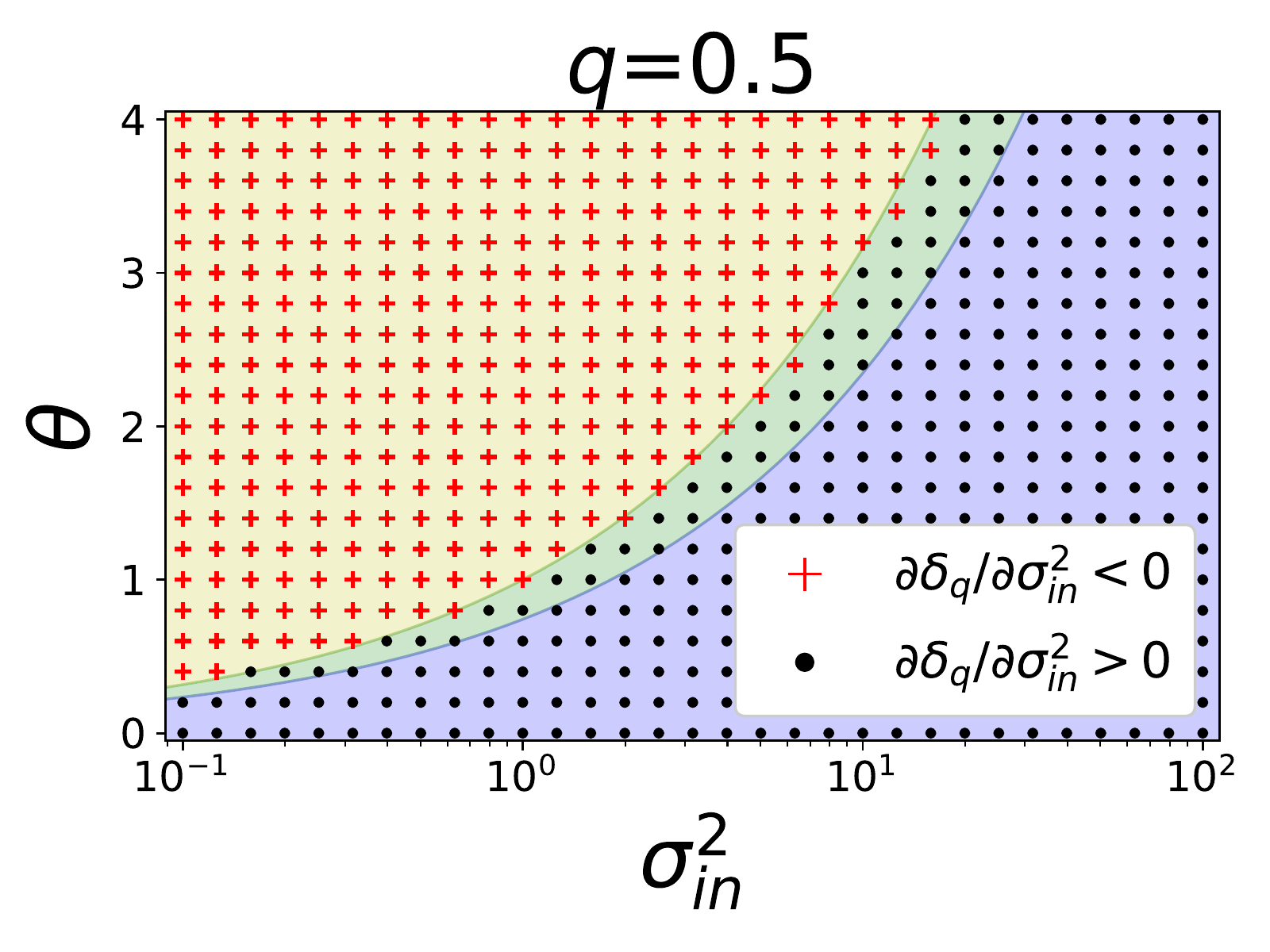}
	\end{minipage}
	 \caption{
	 The case where the robustness function $\delta_{q}(\sigma_{in}^2)$ is defined by the sensitivity $S$. (a), (b) The intrinsic noise dependency of the robustness function $\delta_{q}$. (c), (d) $\sigma_{in}^2$, $\theta$-dependency of $\partial\delta_{q}/\partial\sigma_{in}^2$. (a)--(d) $\partial S/\partial \sigma_{in}^2>0$ (yellow), $\partial S/\partial \sigma_{in}^2<0$ and $\partial^2 S/(\partial \sigma_{in}^2)^2<0$ (green), $\partial S/\partial \sigma_{in}^2<0$ and $\partial^2 S/(\partial \sigma_{in}^2)^2>0$ (blue).
	 }
	\label{fig-sensitivity-robustness}
\end{center}
\end{figure}

In this study, the transmitted information was quantified by the mutual information.
The mutual information is an effective and quantitative measure, even when the input--output relation is nonlinear or when the variance of the output is different for each input.
Therefore, the mutual information is useful when considering complex systems.
However, for the simple systems considered in this study, we can use different measures to evaluate the information transmission between input and output.
In this section, we show that, even if the transmitted information was quantified by some other measure, the robustness against extrinsic noise and the transmitted information could be simultaneously increased by reducing the intrinsic noise when SR appears.
We redefine the measure of the transmitted information as

\begin{align}
	\label{appendix-definition-S}
	S:=\left.\frac{\partial\mathbb{E}[y|x]}{\partial x}\right|_{x=\mu}\mf{.}{}
\end{align}
where $\mathbb{E}[y|x]$ is the mean of the output $y$ given the input $x$, and $\mu$ is the mean of the input $x$. In our models, we set $\mu=0$. This function represents the rate of increase in output with respect to the increase in input, which we refer to as the ``sensitivity.'' 

In the threshold model, from Eqs. (\ref{appendix-p(y=1|x)}) and (\ref{appendix-p(y=0|x)}),

\begin{align}
	\mathbb{E}[y|x]&=\sum_{y\in\{0,1\}}yp(y|x)=\Phi\left(\frac{x-\theta}{\sqrt{\sigma_{ex}^2+\sigma_{in}^2}}\right)
	\label{appendix-E[y|x]}
\end{align}
where $\Phi(x)=\int_{-\infty}^{x}N(z|0,1)dz$ is the cumulative distribution function of the standard normal distribution. From Eq. (\ref{appendix-E[y|x]}), 

\begin{align}
	\frac{\partial \mathbb{E}[y|x]}{\partial x}&=N(\theta|x,\sigma_{in}^2+\sigma_{ex}^2)
	\label{appendix-dE[y|x]dx}
\end{align}
and therefore, from Eq. (\ref{appendix-definition-S}) and $\mu=0$, the analytical solution of the sensitivity in the threshold model is given by

\begin{align}
	S=N(\theta|0,\sigma_{in}^2+\sigma_{ex}^2)\mf{.}{}
	\label{appendix-sensitivity in the threshold model}
\end{align}

When $\theta=0$, the sensitivity $S$ decreases monotonically with the intrinsic noise $\sigma_{in}^2$ (Fig. \ref{fig-sensitivity}(red)); when $\theta>0$, $S$ is maximized with a moderate value of the intrinsic noise $\sigma_{in}^2$ (Fig. \ref{fig-sensitivity}). Therefore, SR appears when the transmitted information is quantified by the sensitivity instead of the mutual information.

We now derive the condition under which SR appears. To simplify the calculation, we define $\sigma^2:=\sigma_{in}^2+\sigma_{ex}^2$. From Eq. (\ref{appendix-sensitivity in the threshold model}), 

\begin{align}
	\frac{\partial S}{\partial \sigma^2}&=\frac{\partial}{\partial \sigma^2}N(\theta|0,\sigma^2)\nonumber\\
	&=\frac{1}{2\sqrt{2\pi}\sigma^5}\left(\theta^2-\sigma^2\right)\exp\left(-\frac{\theta^2}{2\sigma^2}\right)\nonumber\\
	&\begin{cases}
		>0&(\theta>\sigma)\\
		\leq0&(\theta\leq\sigma)\\
	\end{cases}
\end{align}
Therefore, the condition for the appearance of SR is $\theta>0$. Note that this is different from the condition $\theta>2$ in the case where the transmitted information is quantified by the mutual information.

To investigate the robustness against extrinsic noise in terms of the sensitivity, we define the robustness function $\delta_{q}(\sigma_{in}^2)$  as follows:

\begin{align}
	\delta_{q}(\sigma_{in}^2):=\left\{\sigma_{ex}^2\left|\frac{\left.S\right|_{\sigma_{ex}^2=\sigma_{ex}^2}}{\left.S\right|_{\sigma_{ex}^2=0}}=q,\ \ \ q\in(0,1)\right.\right\}
\end{align}
When $\delta_{q}(\sigma_{in}^2)$ is large, the robustness against extrinsic noise $\sigma_{ex}^2$ is high, and vice versa. 

Because  the robustness function $\delta_{q}(\sigma_{in}^2)$ cannot be analytically solved, we derived $\delta_q(\sigma_{in}^2)$ numerically (Fig. \ref{fig-sensitivity-robustness}). When SR does not appear ($\theta=0$), the robustness function $\delta_{q}(\sigma_{in}^2)$ increases monotonically with the intrinsic noise for any $q$ (Fig. \ref{fig-sensitivity-robustness}(a)). As the sensitivity decreases monotonically with any increase in the intrinsic noise, the sensitivity and the robustness against extrinsic noise exhibit a trade-off relationship.  However, when SR appears ($\theta=1$), the robustness function does not always increase with the intrinsic noise (Fig. \ref{fig-sensitivity-robustness}(b)). In particular, when $q$ is close to 1 and the sensitivity is a decreasing and convex upward function with respect to the intrinsic noise, the robustness function decreases with the intrinsic noise (Fig. \ref{fig-sensitivity-robustness}(b)(green)). In this region, by reducing the intrinsic noise, both the sensitivity and the robustness function can be simultaneously increased. Therefore, SR solves the trade-off problem between the transmitted information and the robustness against extrinsic noise. We can see this characteristic over a wider range of parameters (Fig. \ref{fig-sensitivity-robustness}(c), (d)). Therefore, even when the transmitted information is quantified by the sensitivity, the robustness against extrinsic noise and the transmitted information can be simultaneously increased by reducing the intrinsic noise when SR appears. 

\color{black}
\bibliography{graduation_thesis}
\end{document}